\documentclass[useAMS,usenatbib]{mn2e}
\usepackage{graphicx,amsmath,epsfig,rotating,fleqn,multirow}

\newcommand{\figdir}
  {./}
\newlength{\narrowfigurewidth}
\newlength{\figurewidth}
\newlength{\widefigurewidth}
\newcommand{\unit}[1]
  {{\mbox{\rm\,\,#1}}}

\newcommand{\detected}
  {{\rmn{det}}}

\newcommand{\gauss}
  {{\cal{N}}}
\newcommand{\lognorm}
  {{\cal{L}}}
\newcommand{\classstat}
  {{\tt{ClassStat}}}
\newcommand{\mergedclassstat}
  {{\tt{MergedClassStat}}}
\newcommand{\psf}
  {PSF}

\newcommand{\nir}
  {NIR}

\newcommand{\uprime}
  {\mbox{$u$}}
\newcommand{\gprime}
  {\mbox{$g$}}
\newcommand{\rprime}
  {\mbox{$r$}}
\newcommand{\iprime}
  {\mbox{$i$}}
\newcommand{\zprime}
  {\mbox{$z$}}

\newcommand{\ymj}
  {\mbox{$Y$$\!\,-$$J$}}
\newcommand{\ymh}
  {\mbox{$Y$$\!\,-$$H$}}
\newcommand{\ymk}
  {\mbox{$Y$$\!\,-$$K$}}

\newcommand{\hmk}
  {\mbox{$H$$-$$K$}}

\newcommand{\ps}
  {P_{\rmn{s}}}

\newcommand{\etc}
  {etc.}
\newcommand{\etal}
  {et al.}
\newcommand{\eg}
  {e.g.}
\newcommand{\cf}
  {cf.}
\newcommand{\ie}
  {i.e.}
\newcommand{\diff}
  {{\rmn{d}}}
\newcommand{\vect}[1]
  {\mbox{\boldmath ${#1}$}}

\newcommand{\eq}[1]
  {Eq.~(\ref{equation:#1})}

\newcommand{\Eq}[1]
  {Equation~(\ref{equation:#1})}

\newcommand{\sect}[1]
  {Section~\ref{section:#1}}

\newcommand{\tabl}[1]
  {{\mbox Table~\ref{table:#1}}}

\newcommand{\fig}[1]
  {Fig.~\ref{figure:#1}}
\newcommand{\figs}[1]
  {Figs.~\ref{figure:#1}}
\newcommand{\Fig}[1]
  {Figure~\ref{figure:#1}}

\newcommand{\gal}
  {{\rmn{g}}}
\newcommand{\str}
  {{\rmn{s}}}

\newcommand{\pop}
  {{t}}
\newcommand{\npop}
  {{N_{\rmn{\pop}}}}
\newcommand{\popprime}
  {{t^\prime}}

\newcommand{\prob}
  {{\rm{Pr}}}

\newcommand{\weight}
  {W}
\newcommand{\model}
  {\pop}
\renewcommand{\models}
  {\vect{\pop}}

\newcommand{\stat}
  {c}

\newcommand{\ndat}
  {N_{\rmn{d}}}
\newcommand{\nband}
  {N_{\rmn{b}}}

\newcommand{\band}
  {b}
\newcommand{\nparam}
  {N_{\rmn{p}}}
\newcommand{\param}
  {\theta}
\newcommand{\parameters}
  {\vect{\theta}}
\newcommand{\parameter}  
  {\theta}
\newcommand{\data}
  {\vect{d}}
\newcommand{\Data}
  {\vect{D}}
\newcommand{\datum}
  {d}

\newcommand{\density}
  {\rho}

\newcommand{\dvol}
  {\diff \parameter_1 \, \diff \parameter_2 \,
  \ldots \, \diff \parameter_{\nparam}}

\newcommand{\dvolspec}
  {\diff m_1 \, \diff m_2 \,
  \ldots \, \diff m_{\nband} \, \diff \stat}

\newlength{\figwidth}
\newcommand{\stargal}
  {star--galaxy}
\newcommand{\Stargal}
  {Star--galaxy}

\title
  [Bayesian \stargal\ classification]
  {A Bayesian approach to \stargal\ classification}
\author
  [M.\ Henrion \etal]
  {Marc Henrion$^1$\thanks{E-mail: marc.henrion03@ic.ac.uk},
  Daniel J.\ Mortlock$^2$,
  David J.\ Hand$^1$
  and 
  Axel Gandy$^1$
\vspace{7mm}\\
  $^1$Department of Mathematics, 
    Imperial College London, London SW7 2AZ, U.K.\\
  $^2$Astrophysics Group, Imperial College London, Blackett Laboratory,
    Prince Consort Road, London SW7 2AZ, U.K.\\
  }
\begin{document}

\date{Received 2010 ???}

\pagerange{\pageref{firstpage}--\pageref{lastpage}} \pubyear{2010}

\maketitle

\label{firstpage}

\begin{abstract}
\Stargal\ classification is one of the most fundamental 
data-processing tasks in survey astronomy, and a critical
starting point for the scientific exploitation of survey data.
\Stargal\ classification for bright sources can be done with almost complete 
reliability, but for the numerous sources close to a survey's
detection limit each image encodes only limited morphological
information about the source.
In this regime, from which many of the new scientific discoveries are likely to come, 
it is vital to utilise all the available 
information about a source, both from multiple measurements
and also prior knowledge about the star and galaxy populations.
This also makes it clear that it is more useful and realistic
to provide classification probabilities than decisive classifications.
All these desiderata can be met by adopting a Bayesian approach to
\stargal\ classification, and we develop a very general formalism
for doing so.  
An immediate implication of applying Bayes's theorem to this problem 
is that it is formally impossible to combine morphological measurements
in different bands without using colour information as well;
however we develop several approximations that disregard colour 
information as much as possible.
The resultant scheme is applied to data from the 
UKIRT Infrared Deep Sky Survey (UKIDSS), 
and tested by comparing the results to deep 
Sloan Digital Sky Survey (SDSS) Stripe 82 measurements of
the same sources.
The Bayesian classification probabilities 
obtained from the UKIDSS data agree well with the deep SDSS classifications 
both overall (a mismatch rate of $0.022$, compared to 
$0.044$ for the UKIDSS pipeline classifier) 
and close to the UKIDSS detection limit 
(a mismatch rate of $0.068$ compared to 
$0.075$ for the UKIDSS pipeline classifier).
The Bayesian formalism developed here can be applied 
to improve the reliability of any \stargal\ classification 
schemes based on the measured values of morphology statistics alone.
\end{abstract}



\section{Introduction}
\label{section:intro}

Astronomical surveys now gather data on huge numbers of
astronomical objects:
the 2 Micron All Sky Survey (2MASS; \citealt{Skrutskie_etal:2006}),
the Sloan Digital Sky Survey (SDSS; \citealt{York_etal:2000})
and 
the UKIRT Infrared Deep Sky Survey (UKIDSS; \citealt{Lawrence:2006})
have all identified hundreds of millions of distinct sources.
The scale of these projects
immediately necessitates an automated approach to data analysis
(although an intriguing alternative is The Galaxy Zoo project 
  described by \citealt{Lintott_etal:2008}).
Considerable effort has been put into developing 
algorithms which can decompose an image into a smooth background
and a catalogue of discrete objects,
the properties of which must be characterised as well.
Source positions, fluxes and shapes can all
be estimated reliably by using fairly simple moment-based approaches
(\eg, \citealt{Irwin:1985,Bertin_Arnouts:1996}),
but the separation of point-like stars from more extended galaxies
generally requires at least some external astrophysical
information be included.  
As such, the problem of \stargal\ classification 
is well suited to Bayesian methods in which the 
measurements of a given source are combined with 
prior knowledge of the astrophysical populations of
which the source might be a member.
A practical formalism for Bayesian \stargal\ 
classification is developed in this paper.

In \sect{review} the existing methods of \stargal\ classification
are reviewed, with particular emphasis on those which 
are at least partially Bayesian in nature.
A general Bayesian formalism 
for \stargal\ classification is then developed in \sect{prob}, 
and specialised to UKIDSS in \sect{example}.
After analysing a simulated sample in \sect{simanalysis},
the real UKIDSS data are analysed 
-- and the results compared to the classifications from deeper SDSS data --
in \sect{results}.
The relative merits of the Bayesian 
approach to \stargal\ classification are 
summarised in \sect{conc}.

All photometry is given in the native system of the telescope in
question.
Thus SDSS \uprime, \gprime, \rprime, \iprime\ and \zprime\ photometry is 
on the AB
system, whereas UKIDSS $Y$, $J$, $H$ and $K$ photometry is Vega-based.  The 
relevant AB to Vega conversions are given in
\cite{Hewett_etal:2006}.  


\section{Star{\bf{--}}galaxy classification methods}
\label{section:review}

The problem of systematically classifying astronomical 
images as either point-like 
(\ie, generally stars, but also quasars, \etc)
or extended 
(\ie, generally galaxies, but also Galactic nebulae, \etc)
goes back at least as far as \cite{messier:1781}, 
and has been the subject of many investigations in the time since.
This problem is fundamental to astronomy, but 
there is no universally agreed upon method of solving it,
and an almost bewildering number of different approaches have been 
explored.
This is because of varying desiderata 
(\eg, algorithm speed; degree of automation; 
efficiency versus completeness;
the desire for class probabilities versus absolute classification; \etc)
and because different information (morphological and/or colour 
or even spectroscopic) is used.
\cite{Hastie_etal:2008} give a general review of classification methods,
but there is no astronomy-specific equivalent, 
so the various relevant approaches are summarised here.

The starting point for all methods of \stargal\ classification is that
stars and galaxies appear different, 
the latter being more extended (at a given flux level) 
and also exhibiting more variety.
For bright sources these differences are easily distinguished by the human eye 
(as demonstrated so well by the Galaxy Zoo project; 
  \citealt{Lintott_etal:2008});
the challenge is to develop automatic algorithms that can 
perform the same task from measured image properties.
For well-measured, high signal--to--noise ratio sources
that are much brighter than a survey's flux limit, 
\stargal\ separation can be achieved easily, 
and almost any sensible algorithm will achieve the desired results.
The challenge is to treat faint sources correctly,
extracting whatever morphological information is contained in 
the noisy measurements 
whilst also avoiding overly confident 
classification in situations of uncertainty.

The most basic, and probably most commonly used, classification method
is to make simple heuristic cuts in the space of observable image 
properties (and related statistics,
such as the measured second-order moments or kurtosis).
Cuts in this space are either chosen empirically 
(\eg, \citealt{Leauthaud_etal:2007,Kron:1980,Yasuda_etal:2001,Irwin_etal:2010})
or fit to the data
(\eg, \citealt{Macgillivray_etal:1976,Heydon-Dumbleton_etal:1989}). 
Such cut-based methods of \stargal\ separation have a number of benefits:
they are clearly defined; 
they are easy to repeat or simulate;
and they correctly classify the majority of sources.
However cut-based methods also have several important limitations:
the choice of cuts can be essentially arbitrary;
it is difficult to include information about the populations as a whole;
they classify every source with certainty, 
which is almost always unjustified close to the sample's magnitude limits;
and (partly due to the definite classification) 
it is difficult to combine the potentially conflicting 
classifications from different bands or observations.

The arbitrary nature of heuristic cuts can be avoided by using 
automated classification techniques. 
The use of 
neural networks, such as multi-layer perceptrons,
to perform \stargal\ classification was pioneered by 
\cite{Odewahn_etal:1992} and forms a core part of the 
astronomical image analysis packages
{\sc{SExtractor}} \citep{Bertin_Arnouts:1996} 
and 
{\sc{NExt}} \citep{Andreon_etal:2000}.
The use of decision trees has also been explored, 
with both axis-parallel
\citep{Weir_etal:1995a,Ball_etal:2006}
and oblique \citep{Suchkov_etal:2005} trees
applied with varying degrees of success. 
All the above classification methods are objective,
but they are also opaque, 
and it can be hard to predict their behaviour outside
the parameter range in which they were trained and tested.
The need for reliable training data can also be a problem, as 
this can require considerable human input and it is 
difficult to ensure that the necessary parameter range 
is covered.

Any method which decisively classifies all sources has a fundamental problem.
While the images of the bright sources in any sample generally
contain enough information
to justify decisive classifications, many of the faint sources near a survey's
limit should not be classified with such great certainty.
This issue has been tackled using a number of different techniques:
mixture models \citep{Miller_Browning:2003};
fuzzy $k$-means clustering
\citep{Mahonen_Frantti:2000};
semi-supervised clustering
\citep{Jarvis_Tyson:1981};
and difference-boosting networks \citep{Philip_etal:2002}.
These methods are capable of providing non-decisive classifications,
but they still tend towards over-fitting in the absence of 
constraining population models.

The critical point is that, for poorly-measured sources,
there is potentially more information contained in the 
overall constraints on the star and galaxy populations 
than there is in the noisy image of the source in question.
Including both types of information in a logically consistent way
can be achieved by applying Bayes's theorem to 
obtain posterior class probabilities.
Contaminated samples of stars or galaxies could be obtained by adopting 
probability cuts, but ideally the probabilities themselves would 
be retained for all sources.
Even though the source populations are not known perfectly, 
reasonable -- if imprecise -- models should give more realistic results for faint 
sources than any method which does not account for the source populations 
at all.

A fully principled Bayesian formalism for \stargal\ classification 
would involve using (parameterised) models for stars and galaxies 
to evaluate the conditional 
probabilities that a measured image was drawn from 
each of the two populations. 
Comparing these two model likelihoods then yields the 
posterior probability that a source is a star. 
For all its formal correctness, however, 
this is a very involved approach to inferring a 
single number.
Indeed, none of the existing Bayesian implementations of
\stargal\ classification 
(taken to include any method which uses information 
on the source populations as well as the target image) 
have gone to this extreme,
and all adopt a variety of approximations to make the problem 
more tractable.  

Probably the most fully principled Bayesian \stargal\ classification
algorithms implemented to date are those of \cite{Sebok:1979}
and \cite{Bazell_Peng:1998},
who compared fits to the (calibrated) pixel values of the images.
However the need to model, \eg, the spiral arms of brighter galaxies
meant that, paradoxically, 
extra care had to be taken with the brightest
images that should have been easiest to classify.
This is an example of the somewhat counter-intuitive result 
\citep{John:1997} that
attempting to use all the available data does not necessarily 
produce the most discriminating classifier, 
especially when machine learning methods are used
(\eg, \citealt{Bazell_Miller:2005,Ball_etal:2004}).

The problem of galaxy complexity can be overcome by 
using a small number of parameters -- and preferably just one -- 
to characterise how discrepant an image is from those of 
similar stars observed in comparable conditions.
Many morphology statistics have been developed
(\eg, \citealt{Irwin:1985,Scranton_etal:2005}),
and while they are generally not used in a Bayesian context,
any such statistic can be used as a data surrogate.
This fact was utilised very effectively by 
\cite{Scranton_etal:2002},
who used the 
difference between the 
point-spread function (PSF) magnitude
and the best fit galaxy profile model magnitude
(defined as the concentration)
as a measure of the extent of an image.
However,
rather than adopting parameterised 
models of the underlying star and galaxy populations,
they 
fit a mixture model of Gaussians to the double-peaked
\rprime-band
concentration distribution in a number of discrete magnitude ranges.
Overall this combines simplicity and clarity whilst retaining
sufficient information from the image and the populations to
make excellent classifications.
An obvious extension would have been to combine the data from
all five SDSS bands 
(\cf\ \citealt{Koo_Kron:1982,Lupton_etal:2001}).
In general this has proved problematic due to the combination 
of the different depths and the range of source colours 
that can, in particular, result in non-detections 
(\eg, \citealt{Richards_etal:2004}, \citealt{Suchkov_etal:2005} and \citealt{Ball_etal:2006} 
all discard objects that are not detected in all bands).

Multi-band measurements were used in a very different way by 
\cite{Wolf_etal:2001} 
(see also \citealt{Richards_etal:2004}),
who classified sources using colour data.
They utilised
kernel density estimation (KDE) to calculate class densities in  
the space of observable quantities 
(in this case measured colours) and then applied
Bayesian model selection to obtain a final classification. 
The disadvantage of this approach is the 
need for a large training set 
(in order to run the KDE on the stars and galaxies separately).
The use of the noise-convolved,
rather than the intrinsic, distributions
can also result in sub-optimal inferences 
due to the inevitably greater overlap of the observed distributions.

Given the strengths and weaknesses of the various \stargal\
classification methods discussed above, 
we have investigated the utility of a Bayesian approach in which 
the star and galaxy populations are modelled parametrically and 
in which the data from multiple observations can be combined.
The focus is on trying to obtain the best classifications for 
faint objects, with the provision that a decisive answer only be
given if it is merited.


\section{Probabilistic classification of astronomical sources}
\label{section:prob}

Suppose a noisy, seeing-smeared, and pixelated image of a source has been measured.
What can be inferred about the type of object it is?
Assuming there are 
$\npop$ distinct populations of astronomical\footnote{The 
model selection approach followed 
here is conditional on the source being drawn from 
one of the astronomical populations that have explicitly come under
consideration. It 
would also be possible to include 
various non-astronomical noise processes
amongst the models that might explain the data,
such as cosmic rays and random noise spikes.
The difficulty in implementing this idea is that,
whereas most astrophysical populations are at least 
reasonably well constrained, the huge variety of 
poorly understood noise processes are far more difficult
to quantify.}
objects,
$\models = \{ \model_1, \model_2, \ldots, \model_{\npop} \}$, 
under consideration,
the fullest answer to this question is to 
use the available data, 
$\data = \{ \datum_1, \datum_2, \ldots, \datum_{\ndat} \}$, to
calculate the conditional probabilities\footnote{
Throughout this paper we have replaced the more formal 
$\prob (T=\pop | \Data=\data)$, where $T$ is the object type 
variable and $\Data$ is the random vector giving the available data, 
by the less cumbersome, if occasionally ambiguous, 
$\prob (\pop | \data)$.}, 
$\prob (\pop | \data)$,
for each $\pop$. 
Applying Bayes's theorem yields 
\begin{equation}
\label{equation:pssoev}
\prob(\pop | \data)
  = \frac{\prob(\pop) \, \prob(\data | \pop)}
  {\sum_{\popprime = 1}^\npop
    \prob(\popprime) \, \prob(\data | \popprime)} ,
\end{equation}
where
$\prob(\pop)$
is the prior probability that the source is of type $\pop$
and 
$\prob(\data | \pop)$
is the probability (density) of getting the observed data
under the hypothesis that the source is of type $\pop$.
Known as the evidence or the model likelihood, 
the latter is given by 
\begin{equation}
\label{equation:evidence}
\prob(\data | \pop)
  = \int
    \prob(\parameters_\pop | \pop) \,
    \prob(\data | \parameters_\pop, \pop)
    \, \dvol  ,
\end{equation}
where 
$\prob(\parameters_\pop | \pop)$ 
is the usual unit-normalised prior distribution of the 
$\nparam$ model parameters,
$\parameters_\pop 
  = \{ \param_1, \param_2, \ldots, \param_{\nparam} \}$, 
that describe objects of type $\pop$,
and $\prob(\data | \parameters_\pop, \pop)$ 
is the probability (density) of measuring the observed data
given a particular value of this model's parameters 
(\ie, the likelihood).

Whilst \eq{pssoev} is a standard application of Bayes's theorem,
its practical implementation is not so clear in an astronomical context.
Demanding the prior distribution of each population's parameters
be normalised to unity is awkward,
as is the notion of a prior probability of the nature of a source. 
Out of context, the question 
`What is the probability that a source is a star?'
does not have a sensible answer, leaving the priors undefined.
Some constraining information is required, such as a 
range of fluxes or colours, as all probabilities are conditional.
The question `What is the probability that a source
with a magnitude of $i \leq 21.0$ is a star?'
does have a numerical answer, given approximately 
by the observed numbers of stars and galaxies down to the specified limit.
This would yield a reasonable empirical value for the 
priors in \eq{evidence},
although even here the answer depends on Galactic latitude,
due the variation in the stellar density.
The implication is that 
the prior for each population
would have to be defined differently for 
surveys with, \eg, different footprints on the sky or different depths,
a far from satisfactory situation.

These ambiguities can be resolved 
by rewriting \eq{pssoev} as
\begin{equation}
\label{equation:pssoweight}
\prob(\pop | \data)
  = \frac{\weight_\pop(\data)}
    {\sum_{\popprime = 1}^{\npop} \weight_\popprime(\data)},
\end{equation}
where we introduce the weighted evidence,
\begin{equation}
\label{equation:weight}
\weight_\pop(\data)
  = \int
    \density_\pop(\parameters_\pop)
    \prob(\data | \parameters_\pop, \pop)
    \, \dvol.
\end{equation}
Here $\density_\pop(\parameters_\pop)$ 
is the number density 
(per unit solid angle or per unit volume)
of all type $\pop$ sources -- 
not just those that might be detected in the survey under
consideration -- as a function of their parameters\footnote{
In the simple case that $\parameters_\pop$ was a source's apparent
magnitude in a given band, $m$, then $\density_\pop(\parameters_\pop)=
\density_\pop(m)$ would just be the number counts in that band,
but continuing, potentially unbounded, below the detection limit
of the survey in question.
The potentially infinite number of ultra-faint sources is irrelevant
as $\density_\pop(m)$ is multiplied by the likelihood
[$\prob(\hat{m}| m)$ in this simple case] 
which ensures that the product of the source density and the likelihood
is finite and that the integral in \eq{weight} converges.}.
For \eq{pssoweight} to be valid, $\data$ needs to include whether or not the 
source has been detected, as well as its observed properties.

The main benefit of using $\density_\pop(\parameters_\pop)$,
instead of the unit-normalised prior 
$\prob(\parameters_\pop,\pop)=\prob(\pop)\prob(\parameters_\pop|\pop)$,
is that the source density 
has an absolute, empirical 
and context-independent normalisation,
given by the number of observed sources.
Not being dependent on generally arbitrary parameter space
boundaries, it is independent of the details of the current experiment,
and needs only be calculated once. The detection probability is included in $\prob(\data|\parameters_\pop,\pop)$, which is survey-dependent.

Equations~(\ref{equation:weight}) and (\ref{equation:pssoweight}) 
describe a general method for probabilistic classification of 
astronomical sources, 
by explicitly combining 
the information contained in the
measurements of a source 
with existing knowledge
of the populations from which it might have been drawn.
When applied to the more specific problem of \stargal\ classification
these equations simplify further still.


\subsection{Star--galaxy classification}
\label{section:stargalclass}

The probabilistic astronomical classification formalism described above 
can be applied effectively to \stargal\ classification by 
making several simplifying assumptions:
that every source is either a star or a galaxy;
that the useful morphological information in an 
image can be compressed into a single statistic;
and that the source flux is sufficiently well measured that the 
uncertainty in the photometry can be ignored.
Each of these approximations means the 
resultant class probabilities are taken away from the 
ideal value that would be obtained if all the available information 
were utilised, 
but the implicit information loss is only significant to the degree it 
changes the final classifications.
As the bright, well-measured sources in any sample will 
be successfully classified by any sensible algorithm,
it is only necessary to ensure that 
the useful information for the faint sources near the survey limit
is retained.
In the context of \stargal\ separation 
there is no benefit in trying to encode the 
wealth of morphological information present in, \eg,
the image of a bright barred spiral galaxy --
a statistic that accurately represented the degree to which 
a faint source is extended beyond the \psf\ is far more useful.
The guiding principle in the approximations adopted here 
is whether they will significantly alter the classifications of the
ambiguous faint sources.

How many different populations should be considered 
for a typical source detected in an astronomical survey?
The vast majority of known sources are either 
Galactic stars (\ie, $\pop = \str$) or galaxies (\ie, $\pop = \gal$).
The next most common are quasars; but, as their name suggests,
most appear as point--sources in the optical or 
near-infrared (\nir) bands, and 
so can be included with the stars in the context of morphological
classification.
Hence the set of models 
can reasonably be reduced to $\models = \{ \str, \gal \}$.
\Eq{pssoweight} can then be simplified to give the probability 
that a source is a star as 
\begin{equation}
\label{equation:ps}
\ps = \prob(\str | \data) =
  \frac{\weight_\str(\data)}
  {\weight_\str(\data) + \weight_\gal(\data)}.
\end{equation}
Thus the full result of the calculation is just a 
single number, $\ps$.

It is possible to simplify the problem of \stargal\ 
classification by considering only generic measurable 
properties of a source.  
Following the arguments in \sect{review}, 
it is assumed that each of the available images of a source provides only a
single morphology statistic, $\stat$, which encodes
the degree to which it is not point-like.
There is great freedom in how $\stat$ is constructed from the images,
and even what the fiducial stellar value is.
The key point is conceptual: 
the potentially
large data and parameter spaces are both greatly reduced
by the use of a single morphology measure.
The relevant data are simply the measured apparent magnitudes,
$\{ \hat{m}_1, \hat{m}_2, \ldots, \hat{m}_{\nband} \}$,
and measured morphology statistics,
$\{ \hat{\stat}_1, \hat{\stat}_2, \ldots, \hat{\stat}_{\nband} \}$,
in each of the 
$\nband$ bands in which measurements have been made
and in which the source has been detected.
In general it is also necessary to include the fact that the 
source has been detected at all, as this is significantly 
greater for the faintest point-like objects near a survey's detection limit than for extended sources.
Hence the full data vector is 
$\data = \{\detected, \hat{m}_1, \hat{\stat}_1, \hat{m}_2, \hat{\stat}_2, 
  \ldots \hat{m}_{\nband}, \hat{\stat}_{\nband}\}$, where $\detected$ encodes whether 
the source is detected or not. 
The parameters used to describe a source's 
observable intrinsic properties are 
its (true)
apparent magnitudes,
$\{m_1, m_2, \ldots, m_{\nband}\}$, 
in each of the $\nband$ bands,
and its (true) morphology 
statistic\footnote{The notion of a true morphology statistic is somewhat 
artificial,
given that $\stat$ is generally defined in terms of image properties
such as pixel values; however it is taken to be the value of the
morphology statistic that would have been measured if the source 
was observed without photometric noise, but with the smearing of the
observational PSF.
As such $\stat$ is not actually an intrinsic property of the source.
Another potential ambiguity is that 
$\stat$ could have different values in each band,
(\eg, due to star-formation regions in the arms of 
a spiral galaxy being more prominent in shorter wavelength bands), 
although such discrepancies would be strongest
in the better-resolved, brighter galaxies that can be easily classified
anyway.},
$\stat$.
The full parameter vector is then
$\parameters = \{ m_1, m_2, \ldots, m_{\nband}, \stat\}$.

Substituting the above definitions of $\data$ and $\parameters$
into \eq{weight}, the 
weighted evidence can be written as
\begin{equation}
\label{equation:weightexpand}
\weight_\pop(\data)
 = \int
    \density_\pop( m_1, m_2, \ldots, m_{\nband}, \stat )
\end{equation}
\[
\, \mbox{}
  \prob(\detected, \hat{m}_1, \hat{\stat}_1, \hat{m}_2, \hat{\stat}_2, 
    \ldots \hat{m}_{\nband}, \hat{\stat}_{\nband}
    |
    m_1, m_2, \ldots, m_{\nband}, \stat, \pop ) 
\]
\[
\, \mbox{} \dvolspec .
\]
Note that, due to the choice of observable model parameters,
the likelihood now has the same form for both stars and galaxies,
whereas in \eq{evidence} it was population-dependent
(as there was the possibility of using intrinsic physical parameters 
spectral type or Hubble type, which are only defined for 
stars and galaxies, respectively).
The form of the population density and the prior 
can now be treated separately, 
and both can be usefully simplified further.

The likelihood 
should encode photometric uncertainties and
the limitations of the morphological measurements,
as well as correlations between measurements in different bands.
It is, however, reasonable to assume that inter-band 
photometric noise 
correlations are negligible 
(but see \citealt{Scranton_etal:2005}), 
in which case the likelihood becomes a product over the 
$\nband$ bands.
It is also reasonable to assume that the photometric part of the likelihood
is Gaussian in magnitude units -- 
whilst this approximation breaks down for faint sources 
(\eg, \citealt{Mortlock_etal:2010a}), 
all the sources here are unambiguously detected.
It is, however, necessary to include the survey incompleteness,
expressed here as the probability that a source is detected in
at least one band (or, more specifically, in a reference band).
The detection probability is assumed to drop from unity to zero
over a magnitude range $\Delta m_\band$ around the nominal detection 
limit of the survey, $m_{\rmn{lim},\band}$. 
The specific form adopted for the incompleteness is 
\begin{equation}
\label{equation:ErrorFunction}
\prob(\detected | m_\band)
  = 
  \frac{1}{2}\mbox{erfc}
    \left(\frac{m_\band - m_{\rmn{lim},\band}}{\Delta m_\band}\right) ,
\end{equation}
where $\mbox{erfc}(x) 
  = 2 \int_{x}^\infty \gauss(2^{1/2} x^\prime; 0, 1) \, \diff x^\prime - 1$
is the complementary error function, and 
$\gauss(x; \mu, \sigma) 
  = \exp\{-1/2 [(x - \mu) / \sigma]^2\} / [(2 \pi)^{1/2} \sigma]$
is the unit-normalised Gaussian probability density
with mean $\mu$ and variance $\sigma^2$.
Although the detection limits for stars and galaxies are likely to be similar,
the tail of this distribution is significantly longer for stars
(as, being more centrally concentrated, there is a greater
chance of faint stars meeting the detection criteria of most surveys).
A somewhat subtle result of this is that the majority of the very faintest
sources in a sample generated in this way are stars, even for surveys
that are sufficiently deep that 
galaxies are intrinsically much more numerous at such faint fluxes.

Combining the above assumptions, the likelihood for 
stars and galaxies becomes
\begin{equation}
\label{equation:liksep}
\prob(\detected, \hat{m}_1, \hat{\stat}_1, \hat{m}_2, \hat{\stat}_2, 
      \ldots \hat{m}_{\nband}, \hat{\stat}_{\nband}
      |
      m_1, m_2, \ldots, m_{\nband}, \stat, \pop )
\end{equation}
\[
\mbox{}
  = \prod_{\band = 1}^{\nband}
      \gauss \left[\hat{m}_\band ; m_\band, \sigma_\band (m_\band) \right]
      \prob(\hat{\stat}_\band | \stat) ,
\]
where 
$\sigma_\band(m)$ is the magnitude-dependent noise in band $\band$.For the fainter sources of most interest here
(\ie, those within a few magnitudes of the relevant detection limit), 
the noise is background-dominated.
The uncertainty 
for a source
of magnitude $m_\band$ in band $\band$ is then
\begin{equation}
\label{equation:noisemag}
\sigma_\band (m_\band)
  = \frac{1}{5} 10^{2/5(m_\band - m_{{\rmn{lim}}, \band})},
\end{equation}
where 
$m_{{\rmn{lim}}, \band}$ is the limiting magnitude in band $\band$,
at which a source would be detected with,
on average, a signal--to--noise ratio of 5.

The sampling distribution of $\hat{\stat}_\band$ is not as generic as the
distribution of $\hat{m}_\band$ as $\hat{\stat}_\band$ is necessarily a more complicated
statistic, the definition of which is survey-dependent.  
A common choice
(\eg, \citealt{Irwin_etal:2010}) for stars at least, is to define $\stat$ 
such that $\prob(\hat{\stat}_\band | \stat) = 
\gauss(\hat{\stat}_\band ; 0, 1)$
by construction, 
although even in such situations this relationship is not
always satisfied empirically (\cf\ \sect{ClassStatDist}).
Combined with the fact that almost nothing can be said about the form of 
$\prob(\hat{\stat}_\band | \stat)$
in abstract, it is left general for the moment.

The source density $\density_\pop( m_1, m_2, \ldots, m_{\nband}, \stat )$ 
plays several distinct roles in \eq{weightexpand},
most obviously encoding the relative numbers of stars and galaxies 
at a given magnitude, but also 
implicitly including their distribution of colours.
Making this distinction allows the more abstract source density
to be separated into the number counts in a reference band,
$\diff N_\pop / \diff m$,
the conditional distribution of the (true) morphology statistic,
$\prob(c | m, \pop)$,
and 
a conditional magnitude-dependent colour distribution,
$\prob(m_1-m_2, m_2-m_3, \ldots, m_{\nband-1} - m_{\nband} | m)$.
The likelihood
could also be re-written as a function of one reference magnitude
and colour terms $m_1-m_2$, $m_2-m_3$, \etc, without loss of information.
One important implication is 
that it is formally impossible to separate 
colour and morphological information in attempting to perform
\stargal\ separation using multi-band data.
The fact that the morphology statistic of a galaxy depends on its 
magnitude means that some colour-dependent calibration of this 
relationship is required and that this is different for stars and 
galaxies due to their different colours.
From a Bayesian perspective this is very natural:
all the available data (and external information) should be 
brought to bear in any inference problem, 
with any separability falling out as a matter of course.  
However \stargal\ classification is often an intermediate step 
towards a specific science goal, including potentially exploratory
work such as searching for unusual objects.  
In such cases it is often desirable to use colour information alone
(\eg, to search for compact galaxies, as in \citealt{Drinkwater_etal:2003})
or to use morphological information alone 
(\eg, to search for point--sources with unusual colours),
but \eq{weightexpand} shows that the two are inextricably linked. 
Indeed, \cite{Baldry_etal:2010} use morphology jointly with colour 
information to perform the galaxy target selection for the Galaxy And Mass Assembly (GAMA) survey.
It is possible to produce heuristic statistics which depend only on colour
or morphology, but a self-consistent Bayesian approach to \stargal\
classification must include both -- or make 
significant approximations.  

It is the latter approach that is followed here, 
by the potentially extreme step of 
ignoring the uncertainty in the 
measured photometry,
and instead treating a source's measured magnitude in each band,
$\hat{m}_\band$, and its true magnitude, $m_\band$, 
as identical.
This approximation is only justified because of this 
peculiar nature of the problem at hand.  
Given that the colour information is going to be ignored per se,
the only role it will play in the model is to 
allow the morphology statistics of a source to be compared across
bands. 
For example,
the values of $\ps$ calculated 
for two sources of different colours, but with the same
values of $\hat{\stat}_1$ and $\hat{\stat}_2$, in two bands 
could be quite different if only one was bright enough to be 
well classified in a certain band.  
Provided that the typical value of $\stat$ for an object of
type $\pop$ does not vary rapidly with its magnitude, 
it is a reasonable approximation to adopt the average 
colour relationships for each population.  

Applying the above simplifying assumptions to \eq{weightexpand},
we obtain our final general, if approximate, 
expression for the weighted evidence, 
\begin{equation}
\label{equation:weight_final}
W_\pop (\data) = 
\end{equation}
\[
\, \mbox{}  
   \left. \frac{\diff N_\pop}{\diff m} \right|_{m = \hat{m}} 
   \prob(\detected | \hat{m}, \pop)
    \int
    \prob(\stat | m=\hat{m}, \pop)
    \prod_{\band = 1}^{\nband}
    \prob(\hat{\stat}_\band | \stat) \,
    \diff \stat ,
\]
where $\hat{m}$ is the measured magnitude in the reference band
and
$\diff N_\pop / \diff m$ are the differential number counts of type $\pop$
sources in this band.
Note that the photometric data on the source in question only enters 
\eq{weight_final} in the estimate of the number counts 
and the estimate of the true morphology statistic in each band.
The source's measured values of the morphology statistic in each 
band are used, however, entering through the likelihood terms  of the 
form 
$\prob(\hat{\stat}_\band | \stat_\band)$.
Whilst it is impossible to fully escape the link between the measured
shapes and colours of an object, this formalism emphasizes the former
as much as is possible.

Despite the many simplifications that have been made to obtain
\eq{weight_final}, the presence of the survey-specific 
morphology statistic means that a more specific form of 
$\weight_\pop(\data)$ 
can only be obtained in the context of a 
specific survey or data-set.
The variation in image quality and depth, 
combined with the different choices of morphology measure 
mean that the form of $\ps$ that would be obtained
by inserting \eq{weight_final} into \eq{pssoweight}
is our final generic result.


\section{Star--galaxy classification in UKIDSS}
\label{section:example}

The Bayesian approach to \stargal\ classification described 
in \sect{prob} is reasonably general and could be applied to 
generic optical or NIR observations.
However the need for explicit population models means that 
its performance can only be examined in the context of 
specific combination of bands, depths and image quality,
\ie, a particular survey.
For the purpose of exploring our Bayesian approach to
morphological classification we analyse data from the 
multi-band UKIDSS imaging survey (\sect{UKIDSS}),
utilising the overlap with 
the deeper
SDSS Stripe 82 region (\sect{SDSS})
to provide a verification sample.


\subsection{UKIDSS}
\label{section:UKIDSS}

UKIDSS \citep{Lawrence_etal:2007} is 
a suite of five separate \nir\ surveys
using the
Wide Field Camera (WFCAM; \citealt{Casali_etal:2007})
on the United Kingdom Infrared Telescope (UKIRT).
A detailed technical description of the survey is
given by \cite{Dye_etal:2006},
although there have been several improvements in the
time since \citep{Warren_etal:2007}.
In particular, we analyse the 
UKIDSS Large Area Survey (LAS),
which includes imaging 
in the UKIDSS $Y$, $J$, $H$ and $K$ bands 
(defined in \citealt{Hewett_etal:2006}) 
to average depths\footnote{Depths are given in terms of the magnitude
of a point--source that would, on average, be detected with a
signal--to--noise ratio of 5.} of 
$Y \simeq 20.2$,
$J \simeq 19.6$
$H \simeq 18.8$
and
$K \simeq 18.2$
\citep{Dye_etal:2006,Warren_etal:2007}.
The UKIDSS data are obtained from the 
WFCAM Science Archive\footnote{The WSA is located at
{\tt{http://surveys.roe.ac.uk/wsa/}}.} (WSA;
\citealt{Hambly_etal:2008}),
which supplies both images and processed catalogues of 
detected sources.

Aside from basic image parameters 
(\eg, positions, counts, \etc)
these catalogues include a number of derived statistics,
including an extendedness statistic in each band.
The statistic, as defined in \cite{Irwin_etal:2010}, 
is based on the fact that all the unsaturated stars in each field
have the same average curve of growth 
(\ie, fraction of their total flux as a function of angular radius).
This average can be measured empirically, and 
a mismatch statistic calculated for each source.
In a given magnitude range 
the statistic is scaled so that,
for stars, it has 
zero mean, unit variance and is approximately Gaussian distributed;
this scaled mismatch statistic is referred to as \classstat\ in the WSA.
Extended galaxies (and blended pairs of sources)
have positive \classstat\ values, 
whereas most noise sources
(\eg, cosmic rays), being more compact than the PSF,
have negative \classstat\ values.
\classstat\ encodes much of the important
morphological information in even faint images,
and is a superb morphology statistic.
However because it is a statistic based solely on the image data
(\ie, it does not include prior information about a source's nature)
it cannot encode all the information about a source (as distinct from 
the image of it).
Moreover, there is no well-motivated method of combining the \classstat\
values obtained from multiple measurements of a source.
(In UKIDSS there are combined source probabilities and \classstat\ 
values are reported, but these are heuristic in nature, 
and do not retain all the information present in the band-specific
\classstat\ values.)


\subsection{SDSS Stripe 82}
\label{section:SDSS}

The SDSS \citep{York_etal:2000}
has surveyed $\sim 10^4 \unit{deg}^2$ with single observations in
the \uprime, \gprime, \rprime, \iprime\ and \zprime\ bands
\citep{Fukugita_etal:1996},
to depths of $\uprime \simeq 22.0$,
$\gprime \simeq 22.2$,
$\rprime \simeq 22.2$,
$\iprime \simeq 21.3$ and
$\zprime \simeq 20.5$.
The SDSS has also taken repeat measurements in the Stripe 82 region
(covering the right ascension range $\alpha \leq 60$ and $\alpha \geq 300 \unit{deg}$
and declinations of $|\delta| \leq 0.1$),
reaching depths of $\uprime \simeq 23.6$,
$\gprime \simeq 24.5$,
$\rprime \simeq 24.2$,
$\iprime \simeq 23.8$ and
$\zprime \simeq 22.1$.

The SDSS approach to \stargal\ classification is based on the use of 
model magnitudes, each detected source being fit as both a point--source
(\ie, the measured point--spread function)
and a galaxy 
(\ie, a \citealt{Sersic:1963} profile with one of two different exponents).  
The difference between the two different magnitudes,
termed the concentration, $\stat$,
is then used as a morphology statistic
\citep{Yasuda_etal:2001}.
The basic classification is done by designating 
sources with $\stat \leq 0.145$ as stars and sources with 
$\stat > 0.145$ as galaxies.
Whilst this scheme is very effective, it is also important to note
that the classifications of up to a third of sources 
contradict in different bands \citep{Yasuda_etal:2001}.

\begin{figure*}
\begin{center}
\includegraphics[width=\widefigurewidth]{\figdir 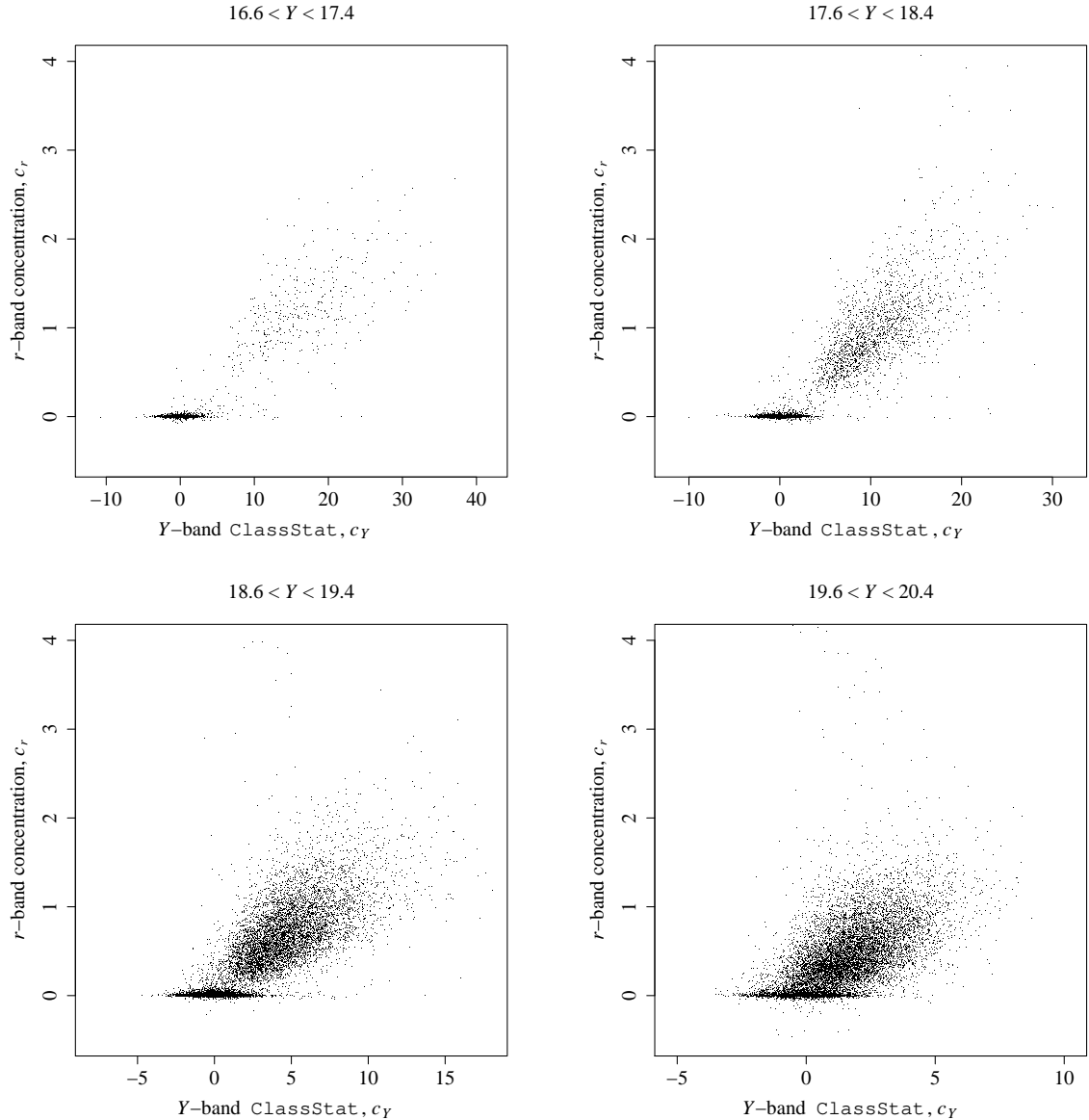}
\caption{SDSS $r$-band concentration plotted against UKIDSS $Y$-band \classstat\ for different magnitude bins.}
\label{figure:CSvsCONC}
\end{center}
\end{figure*}

\begin{figure*}
\begin{center}
\includegraphics[width=\widefigurewidth]{\figdir 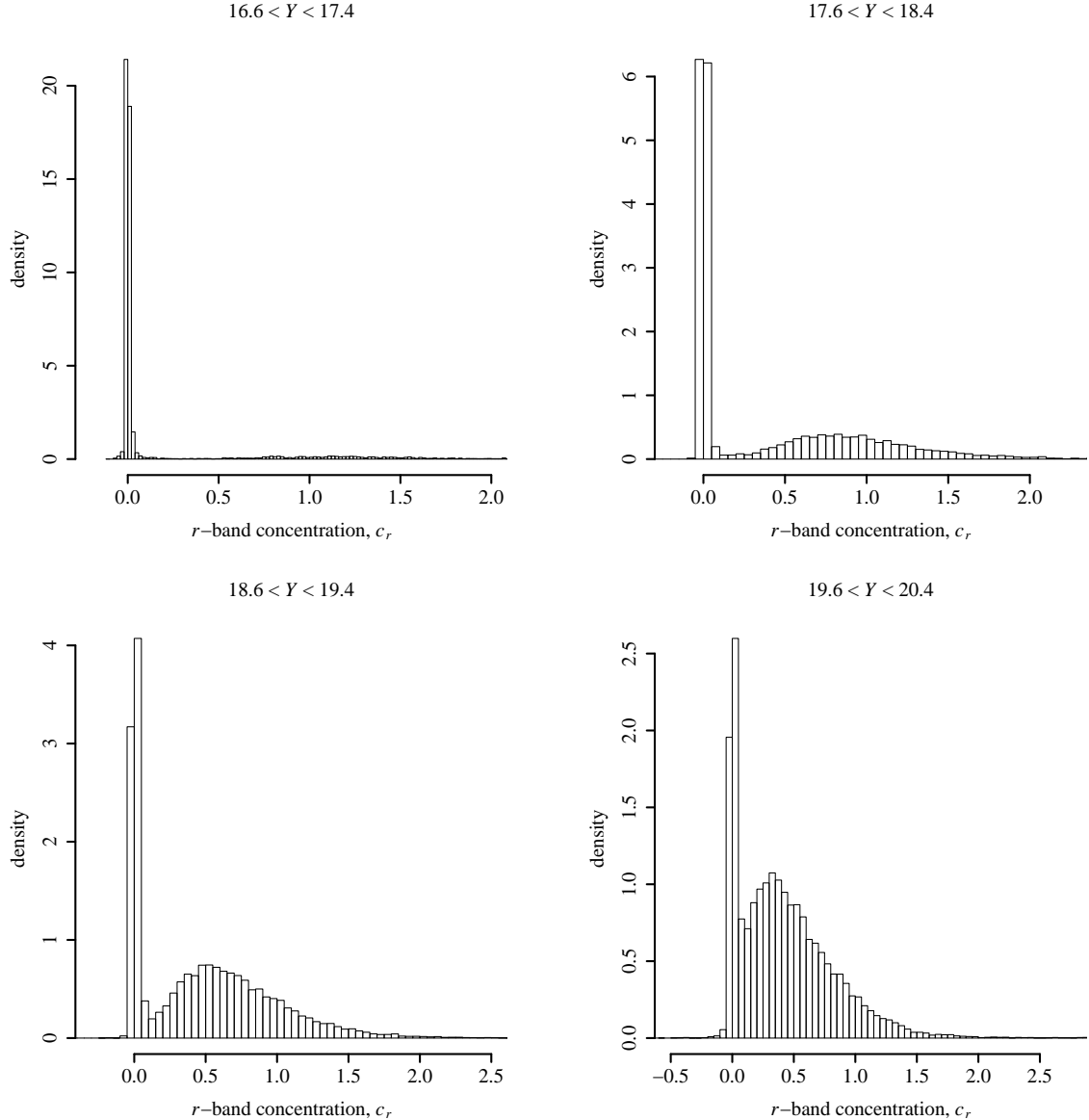}
\caption{One-dimensional slices of SDSS concentration data
  for different magnitude bins.}
\label{figure:1DslicesSDSS}
\end{center}
\end{figure*}

\begin{figure*}
\begin{center}
\includegraphics[width=\widefigurewidth]{\figdir 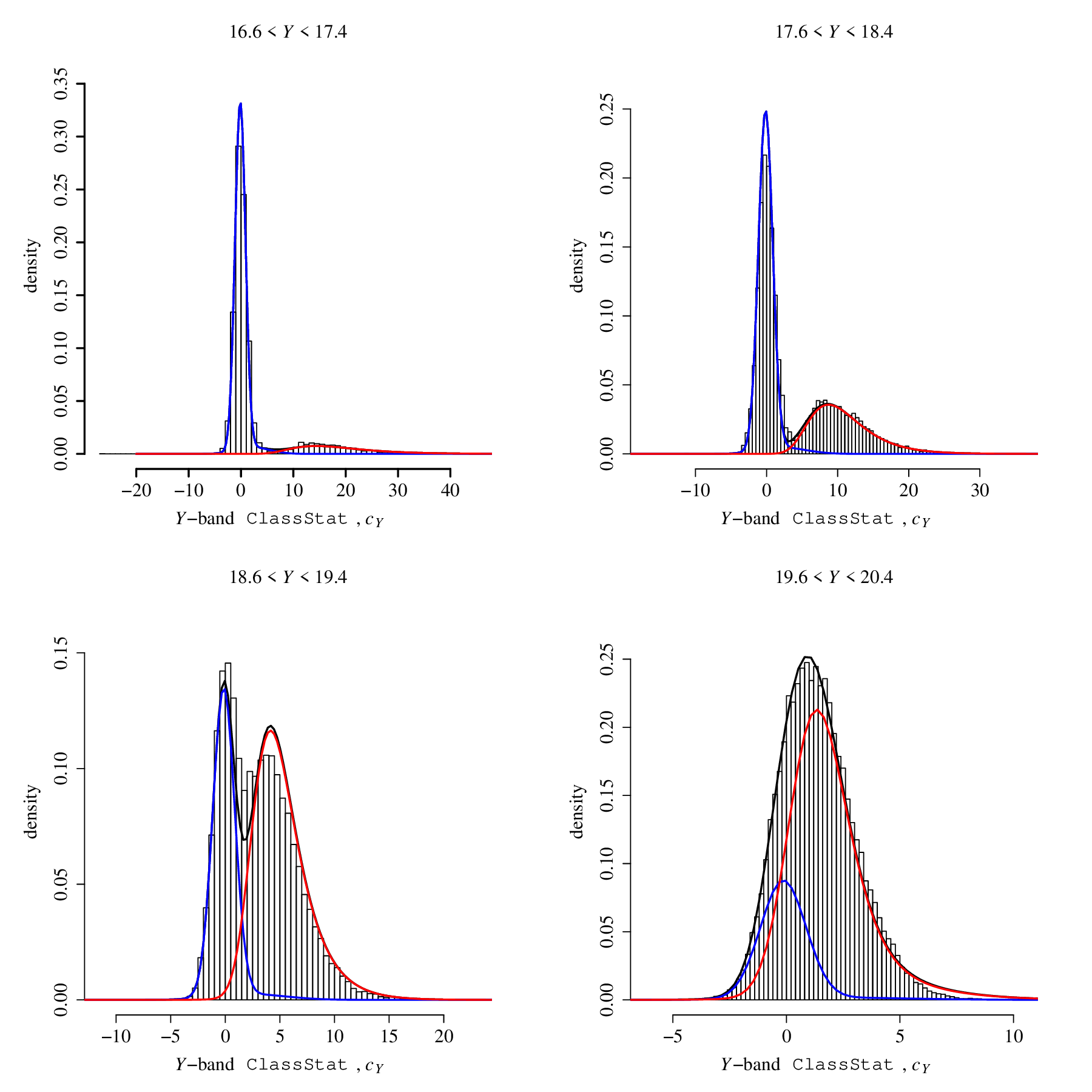}
\caption{One-dimensional slices of UKIDSS \classstat\ data
  for different magnitude bins. Also shown is the fit of our model [overall probability density (black line), star class probability density (blue) and galaxy class probability density (red)], which is discussed in \sect{sim} and further illustrated on \fig{csRealCont}.}
\label{figure:1DslicesUKIDSS}
\end{center}
\end{figure*}

The Stripe 82 data are
significantly deeper than the UKIDSS LAS
(in the sense that all but the reddest sources are detected with
a greater signal--to--noise ratio in Stripe 82 than in the LAS, and an average 
UKIDSS-selected source has $\sigma_r \simeq 0.1\, \sigma_Y$).
Even though the SDSS optical imaging has a significantly larger seeing ($\sim1\,\farcs2$) 
than the UKIDSS NIR data ($\sim0\,\farcs8$), 
the SDSS Stripe 82 data of the
morphologically ambiguous sources near the LAS detection limit 
is able to separate point and extended sources reliably. 
This is illustrated by \figs{CSvsCONC}, \ref{figure:1DslicesSDSS} and \ref{figure:1DslicesUKIDSS}. \fig{CSvsCONC} shows SDSS $r$-band concentration plotted against UKIDSS $Y$-band \classstat. For the faintest two magnitude bins ($Y\simeq19$ and $Y\simeq20$) it is impossible to identify two different populations of sources along the horizontal (\classstat) axis, whereas this is still possible along the vertical (concentration) axis. This is confirmed by the one-dimensional histograms of both classification statistics [\figs{1DslicesSDSS} (concentration) and \ref{figure:1DslicesUKIDSS} (\classstat)]. For $Y\simeq19$, in \fig{1DslicesUKIDSS}, the two populations of sources have almost completely merged, even though the histogram is still bi-modal and for $Y\simeq20$ the two populations of sources cannot be distinguished at all. However the corresponding histograms for SDSS concentration clearly show two distinct populations of sources\footnote{From \fig{PosteriorsOnSDSSConc} it is clear 
that, for $\rprime \ga20.5$ the two clearly distinct populations
of stars and galaxies merge. By limiting ourselves to sources with $16 \leq 
\rprime \leq 20.5$
(thus also avoiding saturated sources) we assume the SDSS class
labels to be correct.}. In particular, for $Y\simeq20$, the SDSS $r$-band class labels  misclassify only $\sim4\%$ of sources (this number is obtained by fitting a Gaussian distribution to the star population and a log-normal to the galaxy population for the SDSS concentration data). This is a very good result when compared to the UKIDSS \classstat\ data which, at this faintness regime, no longer allow a separation into two populations of sources (\fig{1DslicesUKIDSS}).

Hence, for the purpose of \stargal\ separation, we treat the SDSS Stripe 82 
data as definitive classifications against which our Bayesian LAS 
classifications can be tested.

\subsection{Test sample}
\label{section:sample}

Our starting point is a sample of 
$121\,902$ UKIDSS sources in a $14.4 \unit{deg}^2$ area defined by
right ascensions of either $\alpha \leq 60 \unit{deg}$ 
or $\alpha \geq 300 \unit{deg}$ 
and declinations of $|\delta| \leq 0.1$.
This area is entirely within the SDSS Stripe 82 region,
and has been covered by UKIDSS in the $Y$, $J$, $H$ and $K$ bands.
Our main aim is to classify these sources and compare the results to 
the SDSS Stripe 82 classifications.
But to do so requires the preliminary task of generating the 
magnitude-dependent prior distributions of \classstat,
along with the star and galaxy number counts. 
This is not part of the actual classification process
(\ie, it is independent of any single source),
and so is considered separately from the results.


\subsection{Number counts}
\label{section:counts}

The number counts of stars and galaxies provide the prior 
that will be used to classify sources for which the image data are
ambiguous.  
The counts could be obtained from deeper surveys 
(although none exist in all the UKIDSS LAS bands) 
or from physical models of the source populations 
(although this would be unnecessarily complicated).
For the restricted problem of \stargal\ separation, 
however, it is more direct to fit the star and galaxy counts 
to the target sample itself.
At the bright end the numbers are given directly by the data;
at the faint end it is also necessary to assume some weak prior information 
(essentially that a smooth extrapolation from the bright counts is reasonable).

\begin{figure}
\begin{center}
\includegraphics[width=\figurewidth]
  {\figdir 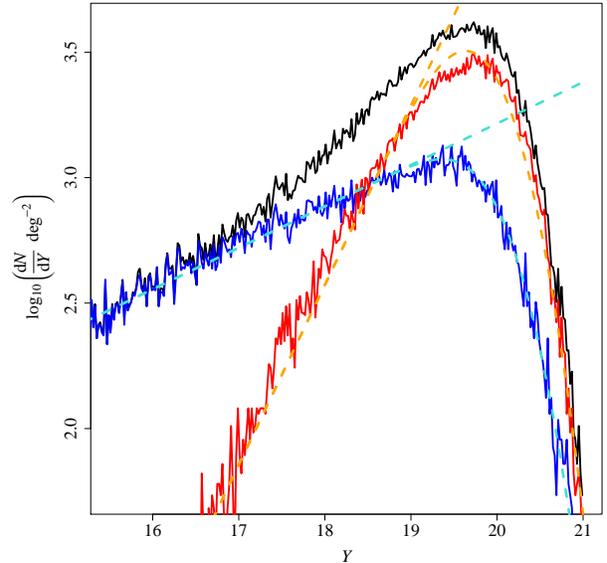}
\caption{Differential number counts of all sources (black), stars (blue) and galaxies (red)
  from UKIDSS observations.
  Classifications are obtained by using our model with number counts obtained by binning the data into equal-sized magnitude bins and fitting simple mixture models to the $\stat_Y$ data in each bin.
  Also shown as dashed lines are the model fits (see \eq{dndy}),
  both with and without a correction for incompleteness.}
\label{figure:NumberCounts}
\end{center}
\end{figure}

For the UKIDSS LAS we
have chosen the $Y$ band as the reference band\footnote{As some sources have
not been observed in all of the bands, for $\hat{m}$ we chose the average of
the magnitudes $\hat{m}_\band$ in the bands in which a given source has
been observed. To convert all of these magnitudes onto the scale of the
reference band we have added the average colours $\ymj,\ymh,\ymk$ to the
magnitudes $\hat{m}_\band$.}.
The observed $Y$ band counts of stars and galaxies 
(identified here by using our model with number counts obtained by binning the data by magnitude and interpolating the parameters)
from the test sample described in \sect{sample} are shown in 
\fig{NumberCounts}.
Both exhibit exponential counts down to $Y \simeq 19$, 
beyond which the survey incompleteness dominates
(as expected, given the average UKIDSS LAS limit of $Y \simeq 20.2$).
For both stars and galaxies the intrinsic number counts are taken
to be of the form
\begin{equation}
\label{equation:dndy}
\density_\pop(Y) = \frac{\diff N_\pop}{\diff Y} = 
  \alpha_\pop \ln(10) N_\pop 10^{\alpha_\pop (Y - Y_0)},
\end{equation}
where $N_\pop$ is the number of sources 
(optionally per unit solid angle, 
although this detail is unimportant as long as 
the same normalising convention is used for stars and galaxies) 
of type $\pop$ brighter than the
reference magnitude $Y_0$, 
and $\alpha_\pop$ is the type-dependent logarithmic slope.
Even though $Y_0$ and $N_\pop$ are degenerate it is convenient
to set $Y_0$ to the $Y$-band magnitude limit, 
in which case $N_\pop$ is approximately equal to the number of 
objects of type $\pop$ in the sample.

In order to fit these parameters, however, it is 
necessary to account for the incompleteness in each band,
denoted here as 
$\prob(\detected | Y)$, 
which was introduced in \eq{ErrorFunction}.
The magnitude limit
$m_{\rmn{lim},\band}$ and incompleteness range $\Delta m_{\band}$ 
are fit in the 
$Y$, $J$, $H$ and $K$ bands for both stars and galaxies.
Fitting ${\diff N_\pop}/{\diff Y} \, \prob(\detected | Y)$  
to the observed UKIDSS counts yields the fits shown
in \fig{NumberCounts}.  
Although there are some discrepancies, the key point is that the
relative numbers of stars and galaxies at a given magnitude 
will give far more accurate prior probabilities 
than, say, an uninformative prior [\ie, $\prob(s) = \prob(g) = 0.5$ for all sources].


\subsection{\classstat\ distributions}
\label{section:ClassStatDist}

\begin{figure}
\includegraphics[width=\figurewidth]
  {\figdir 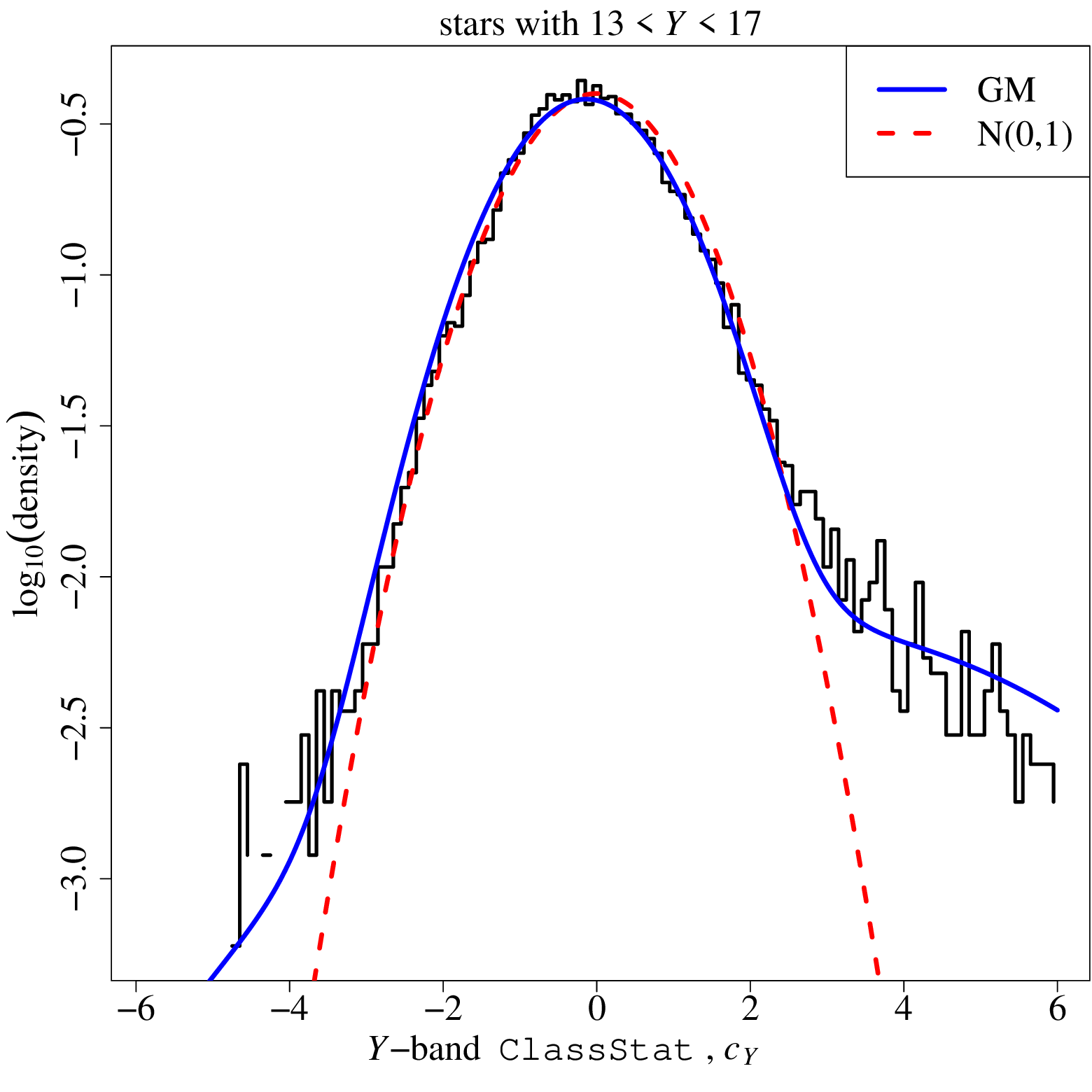}
\caption{The empirical distribution of \classstat\ values of
  bright ($13 \leq Y \leq 17$) UKIDSS sources with $|\stat_Y| \leq 6$ (this selection region is shown on \fig{PostClassProbasSingle}).
  Also shown is a $\gauss(0,1)$ normal distribution and 
  the best-fit Gaussian mixture model defined in \eq{GaussianMixtureModel}.}
\label{figure:csN01vsGM}
\end{figure}

\classstat\ is constructed so that,
on average, $\stat=0$ 
for stars and $\stat>0$ for extended sources. We observe $\hat{\stat}$ however, 
the distribution of which, for isolated stars should be normal
(with zero mean and unit variance), again by construction.
However the observed \classstat\ distribution of 
bright stars (defined as UKIDSS sources with 
$13 \leq Y \leq 17$ and $|\stat_Y|<6$) shown in \fig{csN01vsGM} 
appears to be significantly non-Gaussian.
This impression is confirmed by
the
\cite{Shapiro_Wilk:1965} and one-sample Kolmogorov--Smirnov
\citep{Conover:1999} normality tests.

The distribution of \classstat\ values for the bright stars 
has a slightly negative mean, 
and is weakly positively skewed. 
Due to the positive skewness, using a symmetrical distribution with 
larger tails than a normal (such as Student's $t$-distribution) 
will not result in a good fit.
For the observed \classstat\ distribution we 
have instead adopted a Gaussian mixture model 
of the form 
\begin{equation}
\label{equation:GaussianMixtureModel}
  \prob(\hat{\stat}_\band|\stat)
  = a \gauss(\hat{\stat}_\band-\stat; \mu_1, 1)
    + (1-a) \gauss(\hat{\stat}_\band-\stat; \mu_2, \sigma_2),
\end{equation}
where, for stars, $\stat=0$, and 
$\mu_1$, $\mu_2$ and $\sigma_2$ are free parameters to be fit.
These were fit using a simple maximum likelihood (ML) approach
in each of the four UKIDSS bands. 
The resulting values are given in Table \ref{table:GMparvalsML}, 
and the $Y$ band fit is compared to the data in \fig{csN01vsGM}.

We used the Bayesian information criterion (BIC; \citealt{Schwarz:1978})
to assess the model fit.
As expected, the Gaussian mixture model is a
considerably better fit to the data 
than either fiducial unit-variance Gaussian,
or the Gaussian with ML parameters, resulting in significantly lower BIC values.

The distribution of $\stat$ is more complicated for galaxies
than for stars,
both because galaxies are intrinsically more varied, 
and also because the definition of the morphology statistic
is essentially independent of galaxies' properties.  
For the UKIDSS sample an empirical function was 
sought which could represent the distribution of galaxies' 
$\stat$ values as a function of magnitude.
Particular care was taken to ensure a good fit
close to the survey's limit, for which there is 
minimal morphological information and  
$\stat \rightarrow 0$, even for galaxies.

These desiderata are met by a log-normal distribution:
\begin{equation}
\label{equation:ClassstatGal}
 \prob(\stat | \hat{m}, \pop=\gal) = \lognorm [\stat ; \mu(m),\sigma(m)],
\end{equation}
where 
\begin{equation}
\label{equation:lognormstdpara}
\lognorm(x; \mu, \sigma) = \frac{1}{x\sqrt{2\pi\sigma^2}}
  \exp\left\{-\frac{[\ln (x) - \mu]^2}{2\sigma^2}\right\}.
\end{equation}

Rather than specifying the functions $\mu(m)$ and 
$\sigma(m)$ of the standard parameterisation of the log-normal distribution (Eq.~\ref{equation:lognormstdpara}), 
we have modelled the mean $\mu'(m)$ and standard 
deviation $\sigma'(m)$ of the log-normal distribution\footnote{Here, $\mu'$ and $\sigma'$ are the mean 
and standard deviation of a random variable the logarithm of which is normally 
distributed with mean $\mu$ and standard deviation $\sigma$. These 
parameters are related via 
a standard distributional result: $\mu'=e^{\mu+\sigma^2/2}$ and 
$\sigma'^2=(e^{\sigma^2}-1)\,e^{2\mu+\sigma^2}$.} by the empirical functions below,

\begin{equation}
\label{equation:ClassstatGalMean} 
  \mu'(m) = \left(1-\frac{m}{m_{\rmn{max}}}\right)
\end{equation}
\[
\qquad \qquad \qquad 
  \times\left\{[\nu_1(m-\nu_4)^2+\nu_2(m-\nu_4)+\nu_3]^{\nu_5}+\nu_6\right\} ,
\]
\begin{equation}
\label{equation:ClassstatGalSD} 
  \sigma'(m) = \eta_1 10^{\eta_2(m-11) + 5},
\end{equation}
where $m_{\rmn{max}}$ is the upper detection limit in the reference band and
$\nu_1,\nu_2,\nu_3,\nu_4,\nu_5,\nu_6,\eta_1$ and $\eta_2$ 
are free parameters fitted by a simple least-squares (LS) procedure.

\begin{figure}
\includegraphics[width=\figurewidth]
  {\figdir 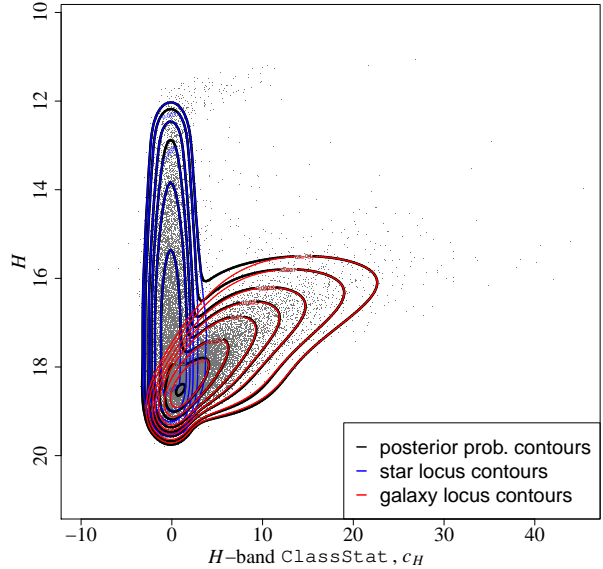}
\caption{The distribution of UKIDSS sources (black points) and the 
  model (contours) in the $H$ band. The case for the $H$ band is plotted as the saturation of bright sources is not as apparent in the $Y$ band. One-dimensional plots of the model fit (this time for the $Y$ band) are shown on \fig{1DslicesUKIDSS}.}
\label{figure:csRealCont}
\end{figure}

The stellar and galactic densities 
implied by our models are shown 
as contours in \fig{csRealCont}, 
along with the sample from which the fit was derived. 
(The $H$ band, rather than the $Y$ band, was chosen as it has the 
highest number of saturated sources, thus emphasizing an aspect of the
data that is not included in the model.)
The fit is not perfect 
(\eg, the true density is underestimated at the bright end and 
slightly overestimated in two regions near the faint end), but is very good.
Also, the bright UKIDSS stars (with $H \la 12.5$) have significantly
positive $\classstat$ values, as they are saturated;
we do not attempt to include this phenomenon as essentially all 
sources bright enough to be saturated in UKIDSS images 
can be classified as stars on the basis of prior information.

\begin{table*}
\begin{center}
\begin{minipage}{110mm}
\caption{Maximum likelihood values, with corresponding standard errors in brackets, for the parameters of the Gaussian mixture model used for the observational noise.}
\begin{tabular}{c|cccc}
 band   & $a$             & $\mu_1$          & $\mu_2$         & $\sigma_2$ \\ \hline
 $Y$    & 0.9453 (0.0030) &  0.1418 (0.0085) & 2.3950 (0.1501) & 3.2021 (0.08600) \\
 $J$    & 0.9436 (0.0053) &  0.1131 (0.0143) & 1.1879 (0.2379) & 3.7199 (0.1776) \\
 $H$    & 0.9601 (0.0033) &  0.1266 (0.0117) & 3.3037 (0.2922) & 3.7523 (0.1617) \\
 $K$    & 0.9474 (0.0039) &  0.0360 (0.0118) & 3.6881 (0.2463) & 3.4449 (0.12975) \\
\end{tabular}
\end{minipage}
\end{center}
\label{table:GMparvalsML}
\end{table*}


\subsection{Simulated data}
\label{section:sim}

Given that the distribution 
of magnitudes and morphology statistics
described above was developed sequentially,
it is important to perform an end--to--end test of the entire model.

The first stage of this was to 
generate a sample of simulated sources from
the model.
The algorithm for doing so can be broken down into several steps:
\begin{itemize}
  \item{Draw a true $Y$ band magnitude from the total 
    (star + galaxy) number count model given in \eq{dndy}.}
  \item{Determine the type (star or galaxy) of the object 
     from the relative number counts at this $Y$ band magnitude.}
  \item{Use the average \ymj, \ymh\ and \ymk\ colours for stars 
    and galaxies (as shown in \fig{PostColColPlot}) to 
    obtain $J$, $H$ and $K$ band magnitudes.}
  \item{Record the object as being detected in each band with 
     probability given by the incompleteness formula in \eq{ErrorFunction}.}
  \item{Add observational (sky) noise to the true magnitudes in all bands
    by sampling from a Gaussian distribution with zero mean and 
    band-dependent standard deviation given by \eq{noisemag}.}
  \item{Generate \classstat\ values for each band by sampling $\stat$ from \eq{ClassstatGal} for galaxies, setting $\stat=0$ for stars and then sampling from the mixture model given in \eq{GaussianMixtureModel}.}
\end{itemize}

\fig{GlobalClassifSimData} shows a sample of data generated by the above procedure. 
Having verified that generating sources from our model 
can accurately mimic the relevant UKIDSS data, 
the model can now be used with confidence as the prior
needed to perform Bayesian \stargal\ classification.


\section{Analysis of simulated UKIDSS data}
\label{section:simanalysis}

A first test of our Bayesian \stargal\ classification method
is to analyse the simulated UKIDSS data described 
in \sect{sim}.  
As the input star and galaxies distributions are known,
the resultant stellar probabilities are,
given the deliberately imposed restrictions on the use of colour information,
optimal.  
In particular, the numbers and properties of the sources which cannot be
classified decisively are of interest, as any real sources with such 
properties will have $\ps \simeq 0.5$.

\begin{figure}
\includegraphics[width=\figurewidth]
  {\figdir 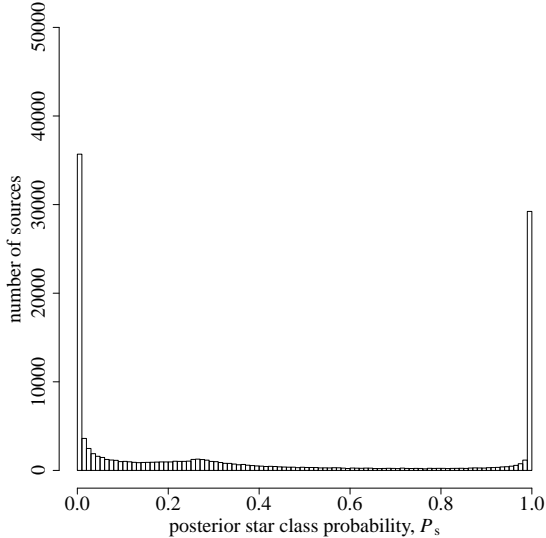}
\caption{Histogram of the posterior star probabilities, $\ps$, evaluated for 
simulated UKIDSS data.}
\label{figure:HistSimData}
\end{figure}

\begin{figure}
\includegraphics[width=\figurewidth]
  {\figdir 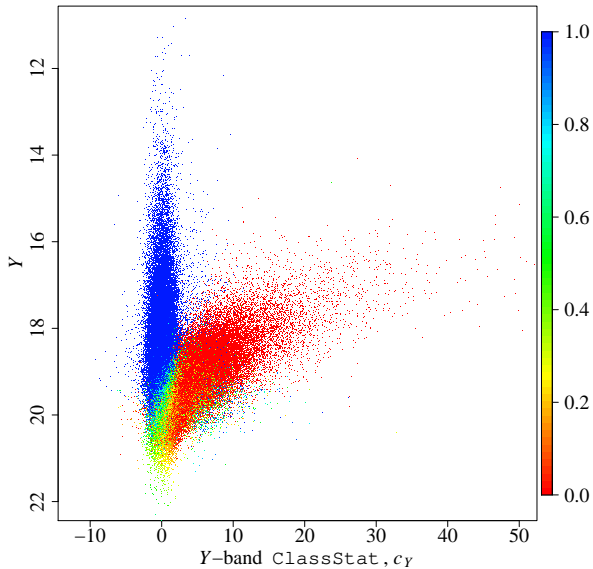}
\caption{Combined star probabilities derived from our Bayesian 
  method for simulated UKIDSS data.}
\label{figure:GlobalClassifSimData}
\end{figure}

\begin{figure*}
\begin{center}
\includegraphics[width=\widefigurewidth]{\figdir 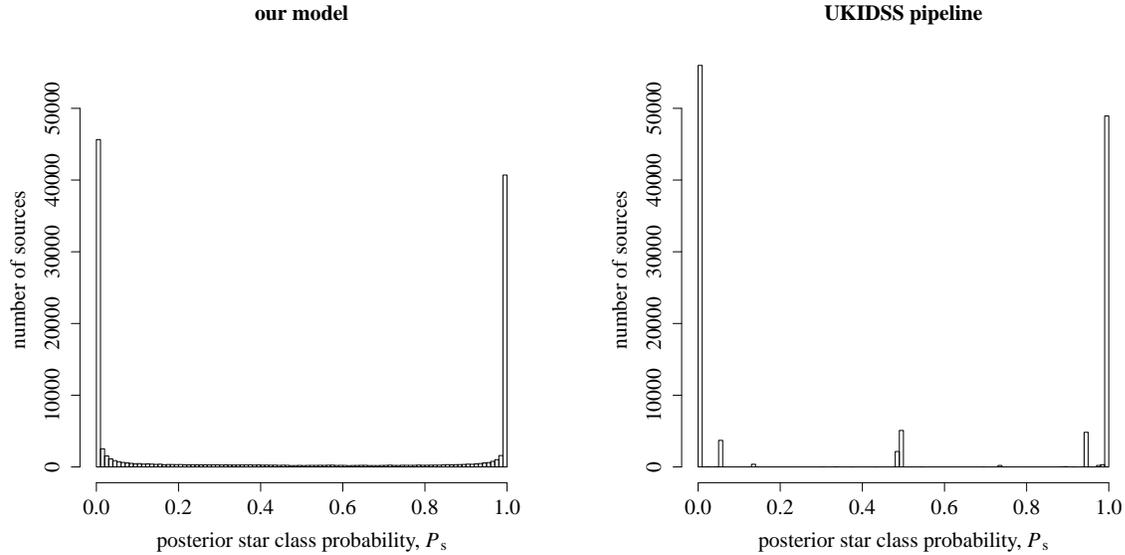}
\caption{Histograms of the posterior star class probability vectors for both our model and the UKIDSS pipeline}
\label{figure:HistPostStarVectors}
\end{center}
\end{figure*}

The distribution of posterior star probabilities for all sources 
is shown in \fig{HistSimData}
and the distribution in 
$Y$ vs.\ $\stat_Y$ space is shown in \fig{GlobalClassifSimData}. 
These results from simulated data can be compared to \figs{HistPostStarVectors} (left) and \ref{figure:PostClassProbasUKIDSSus-pipe} (left), which show the results when our method is applied to real UKIDSS data.
While there is not much difference between \figs{GlobalClassifSimData} and \ref{figure:PostClassProbasUKIDSSus-pipe} (left), there are two noticeable differences between \figs{HistSimData} and \ref{figure:HistPostStarVectors} (left): there are more simulated sources with low star probabilities and there are more sources with $\ps$ clearly different from $0$ and $1$ (\ie, not classified with certainty). In particular there are  many more sources with $\ps\la0.4$, yet clearly non-zero. The former difference can be explained by the fact that there are fewer bright sources (which are predominantly stars and hence have high star probabilities) among the generated data. This means that for equal sample sizes there will be more sources with low star probabilities in the simulated sample when compared to the original data sample.
The increase in sources with less definite classifications is due to the fact that, as acknowledged in Section \ref{section:stargalclass}, our model is not designed to take inter-band photometric noise correlations into account. Thus the simulated data sample contains more sources with seemingly contradicting \classstat\ data in the different bands than a sample of real data.
Both of these differences have only a small effect on the simulated data, and should affect the classification of a negligible number of real sources.


\section{Results}
\label{section:results}

The Bayesian method of \stargal\ classification described above
was applied to the sample of UKIDSS sources in the SDSS Stripe 82 
region, giving single-band star probabilities for every source detected (in each band in which the source was detected), as well as combined probabilities.  
The general properties of the classifier are discussed in 
\sect{Res:PropClassifier}, and then compared to the UKIDSS
classifications (in \sect{Res:UKIDSSpipeline}) and the SDSS classifications 
(in \sect{Res:Stripe82}).


\subsection{Properties of the classifier}
\label{section:Res:PropClassifier}

\begin{table}
\begin{center}
\begin{minipage}{78mm}
\caption{Fraction of sources with posterior probabilities between $0.4$ and $0.6$ for both the single-band models and the joint model. The fractions for the joint model are not the same across the four bands as we only consider sources that are observed in the given bands. So while the probabilities for the joint model are obviously the same across all bands, the fractions in the table above vary across bands as the number of observed sources vary across bands.}
\begin{tabular}{c|rrrr}
  band               & $Y$    & $J$    & $H$    & $K$    \\ \hline
  single-band model  & 0.0332 & 0.0384 & 0.0322 & 0.0336 \\ 
  joint model        & 0.0254 & 0.0254 & 0.0201 & 0.0155 \\
\end{tabular}
\end{minipage}
\end{center}
\label{table:singVSjoint}
\end{table}

\begin{figure}
\begin{center}
\includegraphics[width=\figurewidth]{\figdir 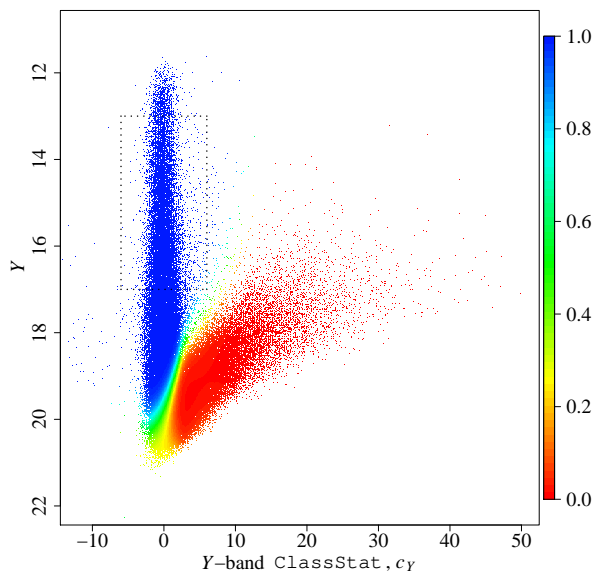}
\caption{Single-band star class probabilities (Y band). The dotted box represents the selection region for the sources from \fig{csN01vsGM}.}
\label{figure:PostClassProbasSingle}
\end{center}
\end{figure}

\Fig{PostClassProbasSingle} shows the single-band posterior star probabilities in $Y-\stat_Y$ space. These can be compared with the probabilities obtained by using the full multi-band model (Fig.~\ref{figure:PostClassProbasUKIDSSus-pipe}). The most notable difference is that for the latter case there seem to be fewer sources which confound the classifier, \ie, with $\ps \simeq 0.5$. Table \ref{table:singVSjoint} lists the fraction of sources for which the classifier gives $0.4 \leq \ps \leq 0.6$. Compared to the single-band model, there is a decrease of at least $25 \unit{per cent}$ in this number for the combined model. While a reduction in the classifier-confounding region is not always desirable, here this decrease translates the fact that the classifier will be at a loss only when the data from different bands are contradictory, or when a source's type is unclear in all the bands in which it was detected. 

\begin{figure}
\begin{center}
\includegraphics[width=\figurewidth]{\figdir 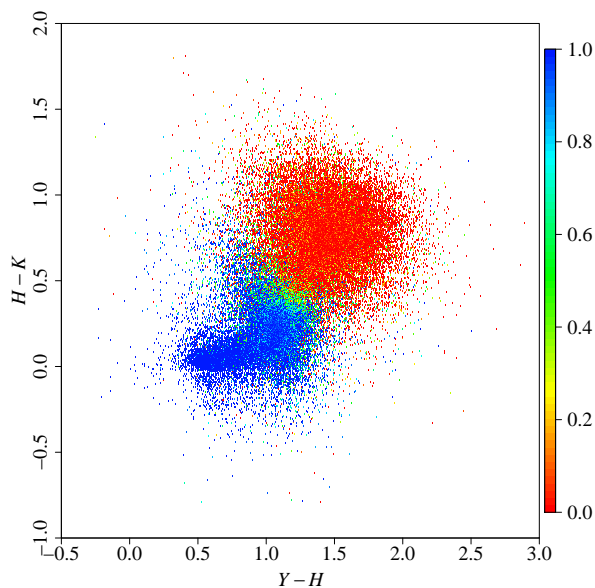}
\caption{Colour--colour plot of the posterior star class probability vector}
\label{figure:PostColColPlot}
\end{center}
\end{figure}

\Fig{PostColColPlot} shows the distribution of the posterior star class probabilities over $\ymh$ vs.\ $\hmk$ space. Even though the model has not been designed to optimise class separation in colour--colour space,
there are two clearly distinct populations.
Furthermore, sources with low star probabilities have 
$\ymh \simeq 1.5$ and $\hmk \simeq 0.8$, as expected.

\subsection{Comparison with UKIDSS pipeline classifications}
\label{section:Res:UKIDSSpipeline}

\begin{figure*}
\begin{center}
\includegraphics[width=\figurewidth]
  {\figdir 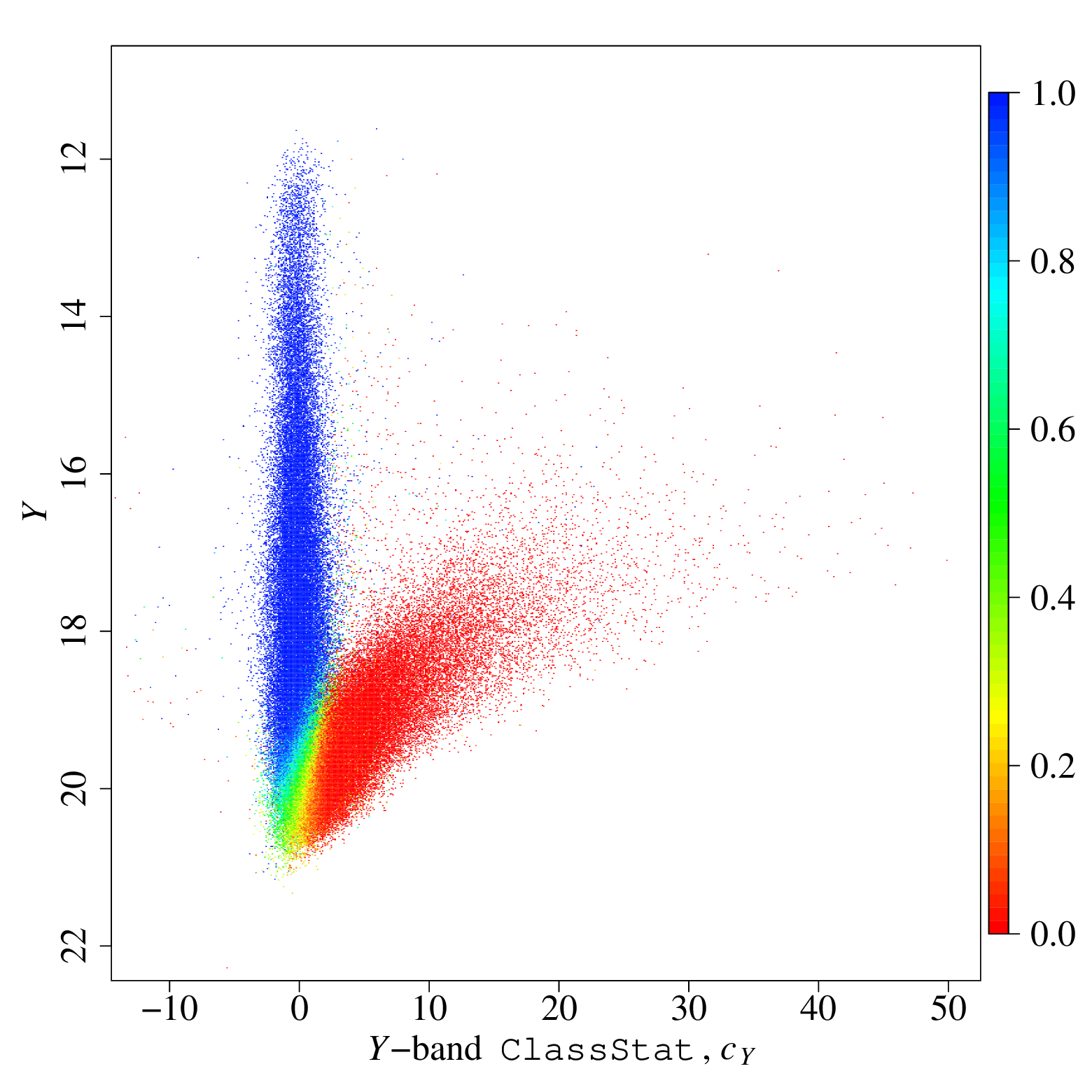}
\includegraphics[width=\figurewidth]
  {\figdir 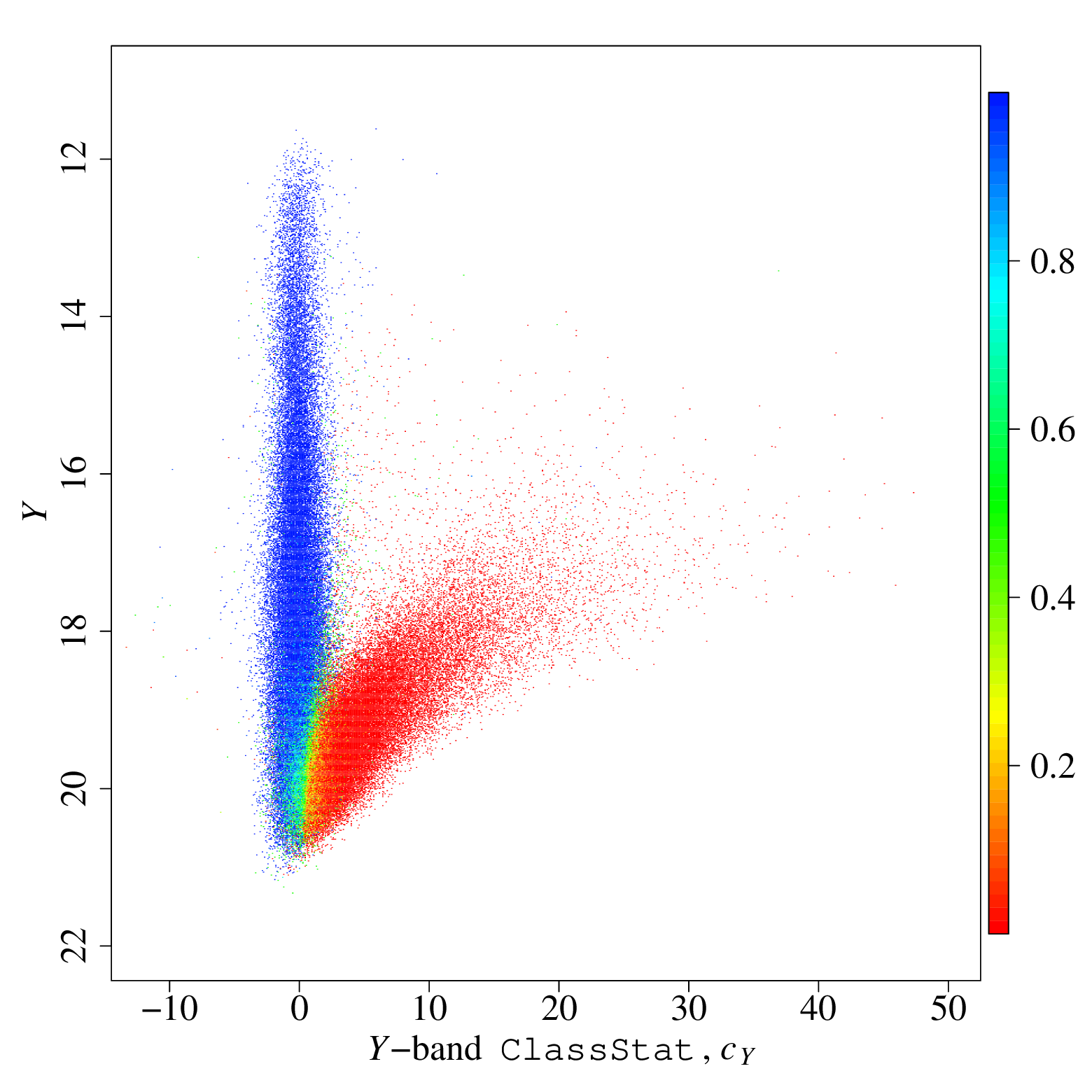}
\caption{Combined star probabilities derived from our Bayesian 
  method (left) and the UKIDSS pipeline (right) as a function 
  of the measured $Y$ band \classstat\ and magnitude.}
\label{figure:PostClassProbasUKIDSSus-pipe}
\end{center}
\end{figure*}

\Fig{PostClassProbasUKIDSSus-pipe} (left) shows the posterior stellar probabilities in the $\stat_Y$ vs.\ $Y$ plane (the choice of band is unimportant, as the $J$, $H$ and $K$ band plots are similar).
It is clear that for the overwhelming majority of objects, in particular those with either $Y \la 18$ or $\stat_Y \ga 5$, the Bayesian classifier gives very definite classifications (\ie, values close to either 0 or 1). Unsurprisingly, the region where the classifier is most often confounded is where the star and galaxy loci merge. 
Indeed, as the two loci overlap completely at the faint end, 
there is very little information regarding object class to be extracted from the measured \classstat\ values, and the prior knowledge drives the classification.

One of the main aims of our classifier is to make the fullest possible
use of whatever morphology statistic is available
-- the UKIDSS \classstat\ statistic in the case considered here --
 and in particular for sources
where it has been measured in multiple bands. 
Several heuristic methods are used to combine multiple measurements 
in the WSA, including simple averaging and a plausible 
-- but again heuristic -- contingency table for sources where the 
\classstat\ measurements in different bands imply contradictory 
classifications.  
Our Bayesian method has the potential to propagate all the information
contained in the individual $\stat$ values correctly, 
albeit at the cost of introducing an explicit -- and 
complicated -- model. 

The UKIDSS pipeline posterior star probabilities can be compared to that from our model (\fig{PostClassProbasUKIDSSus-pipe}). Both classifiers yield similar posterior star probabilities for sources which are fairly bright and/or have large \classstat\ values, but deal differently with faint sources with small \classstat\ values. Apart from a slight shift to the left at the faint end, the UKIDSS pipeline classifier can be seen to consist essentially of a vertical cut on the \classstat\ value.
The classifier-confounding region (\ie, where the classifier outputs probabilities near $0.5$) is fairly small, and, crucially, does not widen at the faint end. Our classifier, however, through the input of prior knowledge, is not limited to taking a vertical cut and the classifier-confounding region is larger, particularly at the faint end. Indeed, near the detection limit, the \classstat\ values carry almost no information concerning object type, as stars and galaxies have similar values at those fluxes. It thus makes very little sense to base a classification on that information. Using prior knowledge is vital for such faint sources.
Our classifier allows a continuous transition from \classstat\ value based classification to prior knowledge based classification. The resulting broader classifier-confounding region is not a drawback: if an object has $\ps\simeq0.5$, it means that, given the observed data, it is impossible to tell whether that source is a star or a galaxy. Artificially coercing posterior classifications to be unambiguous is wrong. If a source cannot be reliably classified, then its posterior probability should reflect this.

Both the posterior probabilities computed by our classifier and the original, observed \classstat\ values can serve as indicators of source type. While one should take a source's flux into account when assessing its \classstat\ data (cf. \fig{PostClassProbasUKIDSSus-pipe}), \classstat\ is designed so as to differentiate between resolved and unresolved sources, and is indeed used to this purpose by the UKIDSS pipeline. Hence it makes sense to compare the posterior class probabilities directly with the \classstat\ values.

\begin{figure*}
\begin{center}
\includegraphics[width=\widefigurewidth]{\figdir 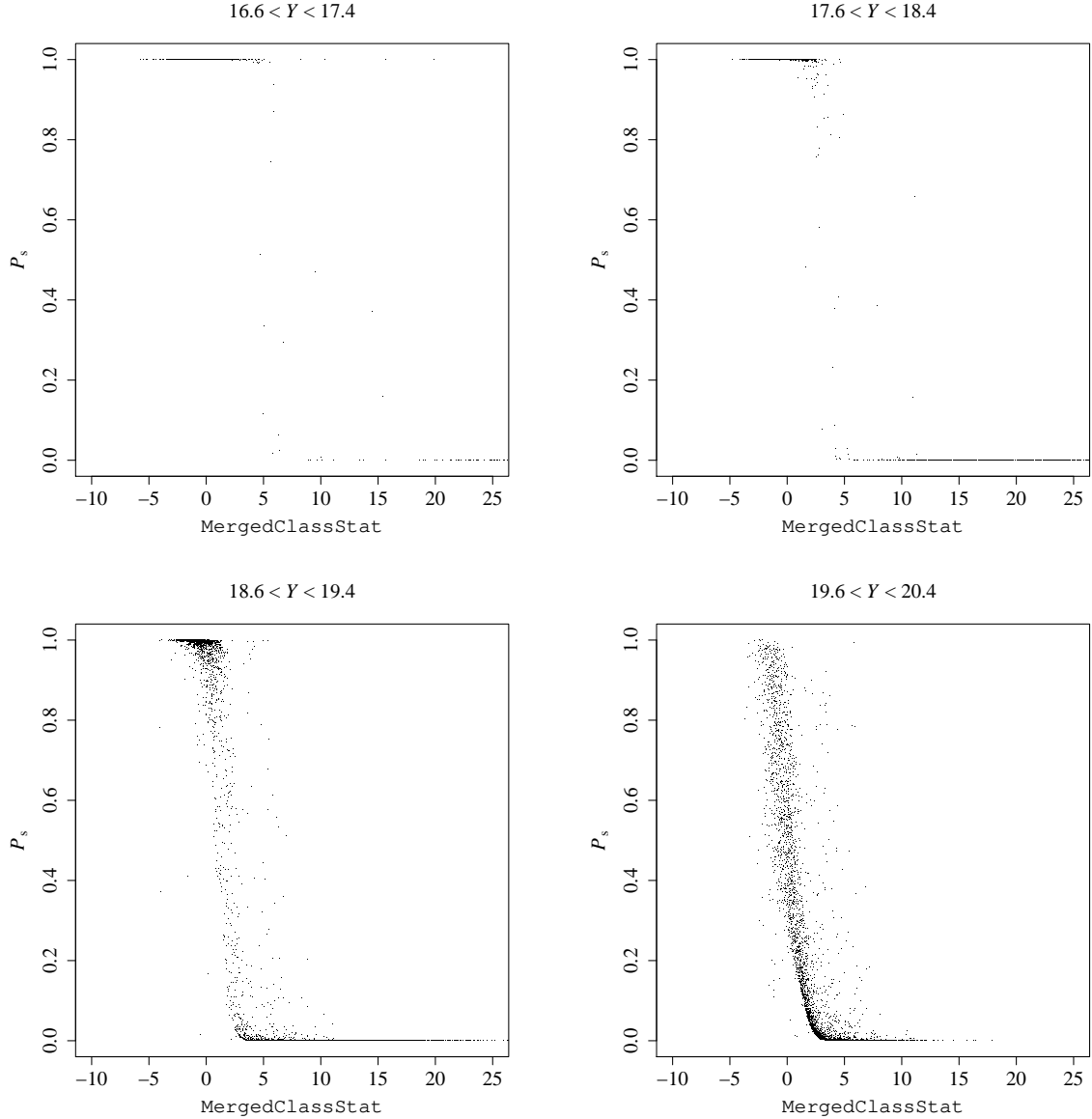}
\caption{posterior star probabilities plotted against \mergedclassstat\
  for different magnitude bins.}
\label{figure:pStarVSmcs}
\end{center}
\end{figure*}

\Fig{pStarVSmcs} summarises the situation for different magnitude regimes. 
At fairly bright magnitudes (\ie, $Y \simeq 17$) most sources have 
$\ps \simeq 1$, except for obviously extended sources with very large \classstat\ values.
At the faint end ($Y \simeq 20$) the classifications are not so 
decisive with few sources having $\ps \simeq 0$ or $\ps \simeq 1$.
The depth of the UKIDSS LAS is such that the surface density of 
stars and galaxies is comparable at the survey's magnitude limit.
This is the most interesting regime for \stargal\ classification
problems: as significantly shallower or deeper surveys would be 
dominated by stars or galaxies, respectively, at their magnitude 
limit, and so essentially all the poorly measured sources would 
be decisively classified purely by the population prior.

However, very low star probabilities ($\ps \la 0.1$) are only reached when \classstat\ exceeds a certain threshold. In the region where star and galaxy populations merge (in magnitude vs.\ \classstat\ space; $Y\simeq19$) a trend is apparent: large \classstat\ values result in low posterior star probabilities. However the reverse is not true: except for sources with extremely low ($\stat_Y<0$) or high ($\stat_Y>10$) \classstat\ values, a source's star probability does not reveal much about its \classstat\ value. In the regions where stars and galaxies are fairly well separated ($Y\simeq17$ and $Y\simeq18$), there is a good correspondence between posterior star probability and ClassStat. 


\subsection{Comparison with SDSS Stripe 82 classifications}
\label{section:Res:Stripe82}

\begin{figure}
\begin{center}
\includegraphics[width=\figurewidth]
  {\figdir 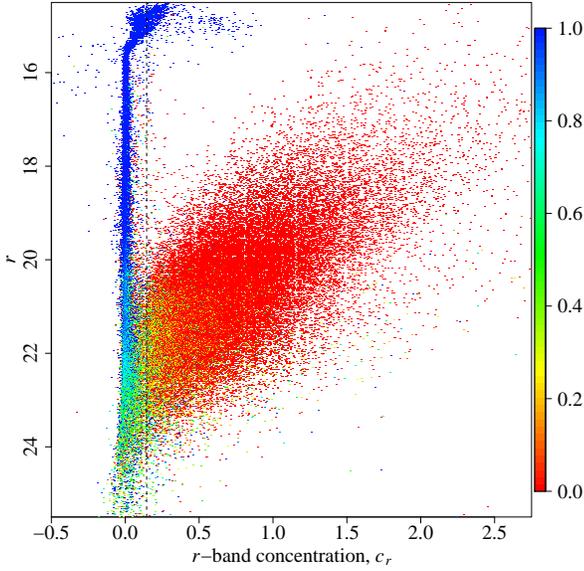}
\caption{UKIDSS posterior star probabilities shown as a function of 
  the measured SDSS Stripe 82 concentration vs.\ $r$-band magnitude. 
  Sources to the left/right of the dotted line (with concentration = 0.145) 
  are classified as stars/galaxies in SDSS.}
\label{figure:PosteriorsOnSDSSConc}
\end{center}
\end{figure}

\Fig{PosteriorsOnSDSSConc} shows the posterior star probabilities from our model as a function of SDSS concentration and  $r$-band magnitude. The dotted line indicates the threshold concentration value ($0.145$) for SDSS star/galaxy labels. Overall there is good agreement with most sources with low $\ps$ lying to the left of the line and sources with high $\ps$ lying to the right.

\begin{figure}
\begin{center}
\includegraphics[width=\figurewidth]{\figdir 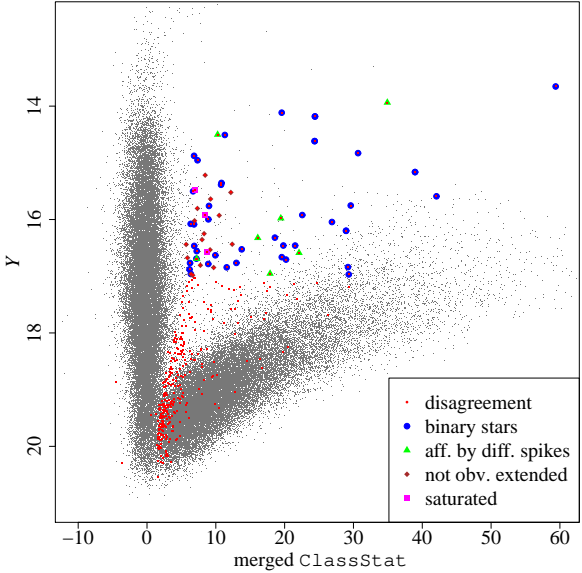}
\caption{The full sample of UKIDSS sources (grey points) with 
inconsistently classified sources highlighted (red).
These are sources with $16<\mbox{average SDSS magnitude}<21.5$
which have either 
$\ps \geq 0.9$ and $\stat_{\rmn{SDSS}} \geq 0.2$ 
or 
$\ps \leq 0.1$ and $\stat_{\rmn{SDSS}} \leq 0.5$.
Most are faint enough that some chance of incorrect classification is
expected on statistical grounds;
an explanation for the brighter sources was sought via visual
inspection, the results of which are indicated.}
\label{figure:SerDis}
\end{center}
\end{figure}

For sources classified with great confidence by both classifiers [\ie, fairly bright, but non-saturated sources ($16\la r\la21.5$) with corresponding UKIDSS posterior star probabilities above $0.9$ and SDSS concentration below $0.05$ or posterior star probabilities below $0.1$ and concentration above $0.2$], we can study those sources for which the two classifiers disagree. \Fig{SerDis} shows that most such sources lie right between the star and galaxy loci.

We have visually checked the sources for which the classifiers disagree. Most are either blended pairs of stars (usually in UKIDSS) or affected by diffraction spikes (in either survey). These sources have been included in \fig{SerDis} and their type is indicated. Sources with large ($\ga15$) \classstat\ values are all either blended binary stars or affected by diffraction spikes of a nearby bright star. 

\begin{figure}
\begin{center}
\includegraphics[width=\figurewidth]{\figdir 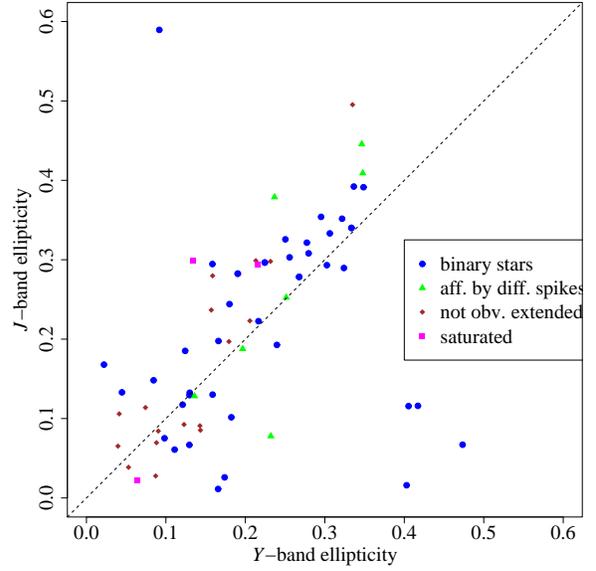}
\caption{$Y$ and $J$ band ellipticities of sources for which both classifiers disagree; the dotted line is the main diagonal}
\label{figure:Ell}
\end{center}
\end{figure}

\Fig{Ell} shows the ellipticities of the misclassified sources, as measured in UKIDSS and SDSS.  In most cases the two measurements are consistent, but for five sources
the estimated ellipticities disagree strongly. Most of the binary stars undetected by UKIDSS, and sources affected by diffraction spikes, lie in the upper-right quadrant of the plot, indicating that UKIDSS indeed detected them as single, extended objects. The five sources far off the diagonal have contradictory data in the different bands. Whether due to noise or inherent source properties, such data will confuse any classifier.

\begin{figure}
\begin{center}
\includegraphics[width=\figurewidth]{\figdir 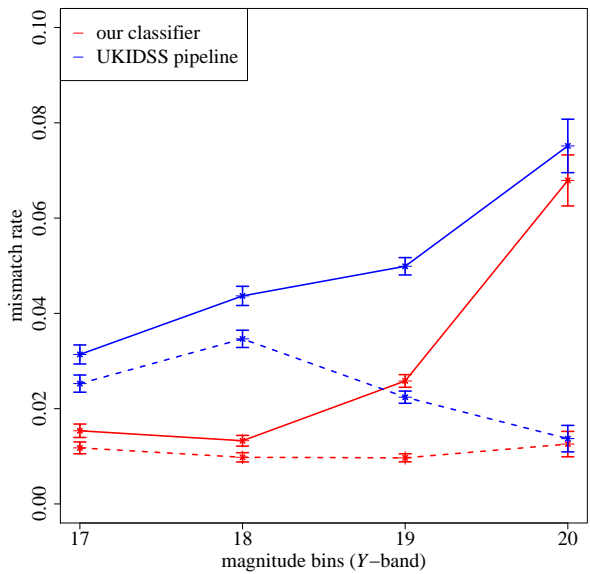}
\caption{Mismatch rates between the SDSS $r$-band class labels and labels based on our classifier (red) and the UKIDSS pipeline (blue). Mismatch rates are shown both for all sources (with $16<r<20.5$; solid lines) and for those sources for which our classifier outputs very definite classifications ($\ps<0.1$ or $\ps>0.9$; dashed lines). The magnitude values on the horizontal axis are the mid-range values of the bins used to compute the rates. Also shown are the standard errors of the mismatch rates.}
\label{figure:mismatchrates}
\end{center}
\end{figure}

Comparing our classifier and the UKIDSS pipeline to the Stripe 82 data, \Fig{mismatchrates} shows the mismatch rates of both classifiers, taking the SDSS $r$-band classifications as a reference. To do this we have converted the posterior probabilities into class labels; an object is labelled as a star if $\ps\geq0.5$, otherwise as a galaxy. We have limited the sources to those with $16<r<20.5$ so as to avoid saturated sources ($r\la16$) and sources for which the uncertainty of the SDSS labels is non-negligible ($r\ga20.5$).  It is clear that the Bayesian classifier is more accurate than the UKIDSS pipeline classifier; even though the difference in performance decreases for fainter magnitudes. For sources with $16.6\leq Y < 17.4$, our classifier achieves a mismatch rate of $0.0154$, compared to $0.0314$ for the UKIDSS pipeline. At the faint end ($Y>20$), the mismatch rates are $0.0679$ (our classifier) and $0.0751$ (UKIDSS pipeline). For all sources with $16<r<20.5$, the mismatch rate for the UKIDSS pipeline ($0.0440$) is more than double that of our classifier ($0.0218$).

\subsection{Value of the Bayesian method}
\label{section:Res:value}

The good performance of both the UKIDSS pipeline classifier and our method 
over the entire sample is unsurprising, 
as most sources are detected with a sufficient signal--to--noise ratio
that they can be classified without effort.
However it is often the case that that the most important sources
scientifically are those close to any new survey's detection limit --
these objects would not have been detected by shallower surveys in 
the same band(s) and inevitably dominate the new discoveries from 
a given data-set.  Hence the inclusion of prior information in the 
Bayesian classifier is most important for just these sources where
it results in significant numbers of more accurate classifications.

Our method provides realistic estimates of the classification uncertainties and allows users, by setting constraints on the posterior classification probabilities $\ps$, to specify the completeness (the fraction of target class sources that have actually been selected) and contamination (the fraction of the selected sources which are not from the target class) of a given selection before starting observations. Thus users can design the selection to suit the survey's aims.

A practical application of our method would be to look at the amount of telescope time that would be required to follow-up a morphologically-selected sample of targets. If one imagines a spectroscopic survey of faint stars, and one was to trust \stargal\ separators such as the ones used by UKIDSS or SDSS versus selecting sources with $\ps>0.9$ from our method, then a certain proportion of telescope time would be spent observing compact / faint galaxies that were misclassified. While there will certainly also be misclassified sources when selecting objects by basing the selection on $\ps$, their proportion can be greatly reduced.

Obviously there is a trade-off between completeness and efficiency when performing source selection. \tabl{wastedtime} lists, for different fluxes, both completeness and efficiency (the fraction of the selected sources which are actually of the target class) for different methods of selecting faint stars, namely selecting sources with $\ps>0.9$ or $\ps>0.5$, using the UKIDSS pipeline single-band or merged class labels, or selecting sources for which the UKIDSS pipeline posterior star probability exceeds $0.9$. While the efficiencies of the different methods are comparable for $Y\simeq17$ and $Y\simeq18$, our method leads to better completeness levels at these fluxes (both for using $\ps>0.5$ and $\ps>0.9$). For $Y\simeq19$ our method with $\ps>0.9$, using the UKIDSS merged class labels and using the UKIDSS pipeline posteriors perform identically. Basing the selection on the UKIDSS $Y$-band class labels or on the posteriors from our method with $\ps>0.5$ leads to higher completeness but lower efficiency (but our method yields a much higher completeness than using the UKIDSS $Y$-band labels and also a marginally larger efficiency). Real differences can, however, be seen at $Y\simeq20$: while our method with $\ps>0.9$ has a much lower completeness level than the UKIDSS pipeline based methods, it also achieves a much higher efficiency. If telescope time is limited and completeness not important, then basing source selection on $\ps$ can lead to a considerable reduction in `wasted' observation time. Using our method with $\ps>0.5$ leads to completeness and efficiency levels more in line with the UKIDSS pipeline based methods.

\begin{table*}
\begin{center}
\begin{minipage}{170mm}
\caption{Completeness (comp.) and efficiency (eff.) for different selection methods at different fluxes. The SDSS Stripe 82 class labels have been taken as reference.}
\begin{tabular}{c||cc|cc|cc|cc}
                                                   & \multicolumn{2}{c}{$16.6<Y<17.4$} & \multicolumn{2}{c}{$17.6<Y<18.4$} & \multicolumn{2}{c}{$18.6<Y<19.4$} & \multicolumn{2}{c}{$19.6<Y<20.4$}          \\
                                                   & comp. & eff.  & comp. & eff.  & comp. & eff.  & comp. & eff.  \\ \hline
 our method with $\ps>0.9$                         & 0.980 & 0.996 & 0.968 & 0.993 & 0.785 & 0.971 & 0.103 & 0.866 \\
 our method with $\ps>0.5$                         & 0.984 & 0.996 & 0.980 & 0.992 & 0.916 & 0.956 & 0.540 & 0.803 \\
 UKIDSS $Y$-band class star label $=-1$ (stars)    & 0.954 & 0.997 & 0.900 & 0.993 & 0.794 & 0.945 & 0.675 & 0.652 \\
 UKIDSS merged class star label $=-1$ (stars)      & 0.964 & 0.997 & 0.922 & 0.993 & 0.782 & 0.972 & 0.626 & 0.799 \\
 UKIDSS pipeline posterior star probability $>0.9$ & 0.963 & 0.997 & 0.921 & 0.993 & 0.782 & 0.972 & 0.626 & 0.799 \\
\end{tabular}
\end{minipage}
\end{center}
\label{table:wastedtime}
\end{table*}


\section{Conclusions}
\label{section:conc}

We have developed a Bayesian formalism for \stargal\ classification 
in optical or \nir\ surveys that combines the morphological
properties of an object (as measured in multiple passbands)
with prior knowledge of the star and galaxy populations.
A fully Bayesian approach must also include colour information
for self-consistency; but, given the aim of combining morphological
information correctly, a number of approximations were 
developed to maximize the influence of the morphological information.

We have demonstrated our method on data from the UKIDSS LAS,
combining morphology statistics measured in the $Y$, $J$, $H$ and $K$
bands (or whatever subset of these a source was detected in).  
The morphology statistic used, \classstat\ \citep{Irwin_etal:2010},
represents a powerful means of data compression from the full image,
and contains almost all the useful information for the faint sources
(which are the main motivation 
for the development of sophisticated \stargal\ classification
techniques).
However, the existing UKIDSS data products include only 
heuristic combinations of the band-specific classifications, 
and the application of the Bayesian method developed here 
makes it possible to extract all the useful UKIDSS information on
a source's morphology in as self-consistent a manner as is possible
without using colour information as well.
In particular, the use of prior information avoids the overly-confident
classification of faint sources, for which the available measurements contain little morphological information. 

Our test sample of UKIDSS LAS sources was chosen to 
lie in the multiply-scanned SDSS Stripe 82 region, 
giving us independent and almost totally reliable classifications 
of all our sources. 
(This is a rare situation outside simulations,
and an opportunity that could be used for a number of similar 
testing schemes.)
Converting the posterior probabilities into class labels, the Bayesian 
classifier achieves an error rate of $0.068$ at the UKIDSS detection 
limit, compared to $0.075$ for the UKIDSS pipeline. For all 
non-saturated sources, the error rate for our model lies at $0.022$, 
compared to $0.044$ for the UKIDSS pipeline.

The Bayesian model used to separate stars and galaxies described here 
can be very easily applied to other surveys with similar statistics measuring 
the extendedness of sources. The multiple advantages of such a classifier 
(posterior probabilities, use of prior knowledge, rigorous computation of 
multi-band classifications, ability to cope with missing detections) and its 
good performance exhibited for the UKIDSS data provide a strong argument 
in favour of a wider use of this methodology. In particular the use of our method 
can improve the efficiency of telescope time.


\section*{Acknowledgments}

The results presented here would not have been possible without the 
efforts of the many people involved in the SDSS and UKIDSS projects.
In particular, we thank Nigel Hambly, Mike Irwin and Steve Warren for help 
in understanding the intricacies of the UKIDSS pipeline.

We also wish to thank the reviewer of the paper who made several insightful comments and helped us to further clarify the text.

\noindent Marc Henrion was supported by an EPSRC research studentship, and David Hand was partially supported by a Royal Society Wolfson Research Merit Award.


\bibliographystyle{mn2e}
\bibliography{references}

\begin{thebibliography}{}

\bibitem[\protect\citeauthoryear{{Andreon}, {Gargiulo}, {Longo}, {Tagliaferri}
  \& {Capuano}}{{Andreon} et~al.}{2000}]{Andreon_etal:2000}
{Andreon} S.,  {Gargiulo} G.,  {Longo} G.,  {Tagliaferri} R.,    {Capuano} N.,
  2000, \mnras, 319, 700

\bibitem[\protect\citeauthoryear{Baldry et~al.,}{Baldry
  et~al.}{2010}]{Baldry_etal:2010}
Baldry I.~K.,  et~al., 2010, \mnras, 404, 86

\bibitem[\protect\citeauthoryear{{Ball}, {Brunner}, {Myers} \& {Tcheng}}{{Ball}
  et~al.}{2006}]{Ball_etal:2006}
{Ball} N.~M.,  {Brunner} R.~J.,  {Myers} A.~D.,    {Tcheng} D.,  2006, \apj,
  650, 497

\bibitem[\protect\citeauthoryear{{Ball}, {Loveday}, {Fukugita}, {Nakamura},
  {Okamura}, {Brinkmann} \& {Brunner}}{{Ball} et~al.}{2004}]{Ball_etal:2004}
{Ball} N.~M.,  {Loveday} J.,  {Fukugita} M.,  {Nakamura} O.,  {Okamura} S.,
  {Brinkmann} J.,    {Brunner} R.~J.,  2004, \mnras, 348, 1038

\bibitem[\protect\citeauthoryear{{Bazell} \& {Miller}}{{Bazell} \&
  {Miller}}{2005}]{Bazell_Miller:2005}
{Bazell} D.,  {Miller} D.~J.,  2005, \apj, 618, 723

\bibitem[\protect\citeauthoryear{{Bazell} \& {Peng}}{{Bazell} \&
  {Peng}}{1998}]{Bazell_Peng:1998}
{Bazell} D.,  {Peng} Y.,  1998, \apjs, 116, 47

\bibitem[\protect\citeauthoryear{{Bertin} \& {Arnouts}}{{Bertin} \&
  {Arnouts}}{1996}]{Bertin_Arnouts:1996}
{Bertin} E.,  {Arnouts} S.,  1996, \aaps, 117, 393

\bibitem[\protect\citeauthoryear{{Casali} et~al.,}{{Casali}
  et~al.}{2007}]{Casali_etal:2007}
{Casali} M.,  et~al., 2007, \aap, 467, 777

\bibitem[\protect\citeauthoryear{Conover}{Conover}{1999}]{Conover:1999}
Conover W.,  1999, Practical Nonparametric Statistics, 3rd edn.
John Wiley \& Sons

\bibitem[\protect\citeauthoryear{{Drinkwater}, {Gregg}, {Hilker}, {Bekki},
  {Couch}, {Ferguson}, {Jones} \& {Phillipps}}{{Drinkwater}
  et~al.}{2003}]{Drinkwater_etal:2003}
{Drinkwater} M.~J.,  {Gregg} M.~D.,  {Hilker} M.,  {Bekki} K.,  {Couch} W.~J.,
  {Ferguson} H.~C.,  {Jones} J.~B.,    {Phillipps} S.,  2003, \nat, 423, 519

\bibitem[\protect\citeauthoryear{{Dye} et~al.,}{{Dye}
  et~al.}{2006}]{Dye_etal:2006}
{Dye} S.,  et~al., 2006, \mnras, 372, 1227

\bibitem[\protect\citeauthoryear{{Fukugita}, {Ichikawa}, {Gunn}, {Doi},
  {Shimasaku} \& {Schneider}}{{Fukugita} et~al.}{1996}]{Fukugita_etal:1996}
{Fukugita} M.,  {Ichikawa} T.,  {Gunn} J.~E.,  {Doi} M.,  {Shimasaku} K.,
  {Schneider} D.~P.,  1996, AJ, 111, 1748

\bibitem[\protect\citeauthoryear{{Hambly} et~al.,}{{Hambly}
  et~al.}{2008}]{Hambly_etal:2008}
{Hambly} N.~C.,  et~al., 2008, \mnras, 384, 637

\bibitem[\protect\citeauthoryear{Hastie, Tibshirani \& Friedman}{Hastie
  et~al.}{2008}]{Hastie_etal:2008}
Hastie T.,  Tibshirani R.,    Friedman J.,  2008, The Elements of Statistical
  Learning: Data Mining, Inference, and Prediction, 2nd edn.
Springer-Verlag

\bibitem[\protect\citeauthoryear{{Hewett}, {Warren}, {Leggett} \&
  {Hodgkin}}{{Hewett} et~al.}{2006}]{Hewett_etal:2006}
{Hewett} P.~C.,  {Warren} S.~J.,  {Leggett} S.~K.,    {Hodgkin} S.~T.,  2006,
  MNRAS, 367, 454

\bibitem[\protect\citeauthoryear{{Heydon-Dumbleton}, {Collins} \&
  {MacGillivray}}{{Heydon-Dumbleton} et~al.}{1989}]{Heydon-Dumbleton_etal:1989}
{Heydon-Dumbleton} N.~H.,  {Collins} C.~A.,    {MacGillivray} H.~T.,  1989,
  \mnras, 238, 379

\bibitem[\protect\citeauthoryear{{Irwin}}{{Irwin}}{1985}]{Irwin:1985}
{Irwin} M.~J.,  1985, \mnras, 214, 575

\bibitem[\protect\citeauthoryear{{Irwin}, {Lewis}, {Hodgkin} \&
  {Gonzales--Solares}}{{Irwin} et~al.}{2010}]{Irwin_etal:2010}
{Irwin} M.~J.,  {Lewis} J.,  {Hodgkin} S.~T.,    {Gonzales--Solares} 2010,
  MNRAS, in preparation

\bibitem[\protect\citeauthoryear{{Jarvis} \& {Tyson}}{{Jarvis} \&
  {Tyson}}{1981}]{Jarvis_Tyson:1981}
{Jarvis} J.~F.,  {Tyson} J.~A.,  1981, \aj, 86, 476

\bibitem[\protect\citeauthoryear{John}{John}{1997}]{John:1997}
John G.~H.,  1997, PhD thesis, Stanford University

\bibitem[\protect\citeauthoryear{{Koo} \& {Kron}}{{Koo} \&
  {Kron}}{1982}]{Koo_Kron:1982}
{Koo} D.~C.,  {Kron} R.~G.,  1982, A\&A, 105, 107

\bibitem[\protect\citeauthoryear{Kron}{Kron}{1980}]{Kron:1980}
Kron R.,  1980, ApJS, 43, 305

\bibitem[\protect\citeauthoryear{Lawrence et~al.,}{Lawrence
  et~al.}{2007}]{Lawrence:2006}
Lawrence A.,  et~al., 2007, \mnras, 379, 1599

\bibitem[\protect\citeauthoryear{{Lawrence} et~al.,}{{Lawrence}
  et~al.}{2007}]{Lawrence_etal:2007}
{Lawrence} A.,  et~al., 2007, \mnras, 379, 1599

\bibitem[\protect\citeauthoryear{{Leauthaud} et~al.,}{{Leauthaud}
  et~al.}{2007}]{Leauthaud_etal:2007}
{Leauthaud} A.,  et~al., 2007, \apjs, 172, 219

\bibitem[\protect\citeauthoryear{{Lintott} et~al.,}{{Lintott}
  et~al.}{2008}]{Lintott_etal:2008}
{Lintott} C.~J.,  et~al., 2008, \mnras, 389, 1179

\bibitem[\protect\citeauthoryear{Lupton, Gunn, Ivezi\'{c} \& Knapp}{Lupton
  et~al.}{2001}]{Lupton_etal:2001}
Lupton R.,  Gunn J.~E.,  Ivezi\'{c} Z.,    Knapp G.~R.,  2001, in preparation,
  0, 1

\bibitem[\protect\citeauthoryear{MacGillivray, Martin, Pratt, Reddish, Seddon,
  Alexander, Walker \& Williams}{MacGillivray
  et~al.}{1976}]{Macgillivray_etal:1976}
MacGillivray H.~T.,  Martin R.,  Pratt N.,  Reddish V.,  Seddon H.,  Alexander
  L.,  Walker G.,    Williams P.,  1976, \mnras, 176, 265

\bibitem[\protect\citeauthoryear{{M{\"a}h{\"o}nen} \&
  {Frantti}}{{M{\"a}h{\"o}nen} \& {Frantti}}{2000}]{Mahonen_Frantti:2000}
{M{\"a}h{\"o}nen} P.,  {Frantti} T.,  2000, \apj, 541, 261

\bibitem[\protect\citeauthoryear{Messier}{Messier}{1781}]{messier:1781}
Messier C.,  1781, Connaissance des Temps, Pour l'Ann\'{e}e Commune 1783..
pp 227--267

\bibitem[\protect\citeauthoryear{Miller \& Browning}{Miller \&
  Browning}{2003}]{Miller_Browning:2003}
Miller D.~J.,  Browning J.,  2003, Transactions on Pattern Analysis and Machine
  Intelligence, 25, 1468

\bibitem[\protect\citeauthoryear{{Mortlock}, {Patel}, {Warren}, {Hewett},
  {Venemans} \& {McMahon}}{{Mortlock} et~al.}{2010}]{Mortlock_etal:2010a}
{Mortlock} D.~J.,  {Patel} M.,  {Warren} S.~J.,  {Hewett} P.~C.,  {Venemans}
  B.~P.,    {McMahon} R.~G.,  2010, MNRAS, submitted

\bibitem[\protect\citeauthoryear{{Odewahn}, {Stockwell}, {Pennington},
  {Humphreys} \& {Zumach}}{{Odewahn} et~al.}{1992}]{Odewahn_etal:1992}
{Odewahn} S.~C.,  {Stockwell} E.~B.,  {Pennington} R.~L.,  {Humphreys} R.~M.,
   {Zumach} W.~A.,  1992, \aj, 103, 318

\bibitem[\protect\citeauthoryear{{Philip}, {Wadadekar}, {Kembhavi} \&
  {Joseph}}{{Philip} et~al.}{2002}]{Philip_etal:2002}
{Philip} N.~S.,  {Wadadekar} Y.,  {Kembhavi} A.,    {Joseph} K.~B.,  2002,
  \aap, 385, 1119

\bibitem[\protect\citeauthoryear{{Richards} et~al.,}{{Richards}
  et~al.}{2004}]{Richards_etal:2004}
{Richards} G.~T.,  et~al., 2004, ApJS, 155, 257

\bibitem[\protect\citeauthoryear{Schwarz}{Schwarz}{1978}]{Schwarz:1978}
Schwarz G.,  1978, The Annals of Statistics, 6, 461

\bibitem[\protect\citeauthoryear{{Scranton}, {Connolly}, {Szalay}, {Lupton},
  {Johnston}, {Budavari}, {Brinkman} \& {Fukugita}}{{Scranton}
  et~al.}{2005}]{Scranton_etal:2005}
{Scranton} R.,  {Connolly} A.~J.,  {Szalay} A.~S.,  {Lupton} R.~H.,  {Johnston}
  D.,  {Budavari} T.,  {Brinkman} J.,    {Fukugita} M.,  2005, \aj, submitted

\bibitem[\protect\citeauthoryear{{Scranton} et~al.,}{{Scranton}
  et~al.}{2002}]{Scranton_etal:2002}
{Scranton} R.,  et~al., 2002, \apj, 579, 48

\bibitem[\protect\citeauthoryear{{Sebok}}{{Sebok}}{1979}]{Sebok:1979}
{Sebok} W.~L.,  1979, \aj, 84, 1526

\bibitem[\protect\citeauthoryear{{S{\'e}rsic}}{{S{\'e}rsic}}{1963}]{Sersic:196%
3}
{S{\'e}rsic} J.~L.,  1963, Boletin de la Asociacion Argentina de Astronomia La
  Plata Argentina, 6, 41

\bibitem[\protect\citeauthoryear{Shapiro \& Wilk}{Shapiro \&
  Wilk}{1965}]{Shapiro_Wilk:1965}
Shapiro S.,  Wilk M.,  1965, Biometrika, 52, 591

\bibitem[\protect\citeauthoryear{{Skrutskie} et~al.,}{{Skrutskie}
  et~al.}{2006}]{Skrutskie_etal:2006}
{Skrutskie} M.~F.,  et~al., 2006, \aj, 131, 1163

\bibitem[\protect\citeauthoryear{{Suchkov}, {Hanisch} \& {Margon}}{{Suchkov}
  et~al.}{2005}]{Suchkov_etal:2005}
{Suchkov} A.~A.,  {Hanisch} R.~J.,    {Margon} B.,  2005, \aj, 130, 2439

\bibitem[\protect\citeauthoryear{{Warren} et~al.,}{{Warren}
  et~al.}{2007}]{Warren_etal:2007}
{Warren} S.~J.,  et~al., 2007, \mnras, 375, 213

\bibitem[\protect\citeauthoryear{{Weir}, {Fayyad} \& {Djorgovski}}{{Weir}
  et~al.}{1995}]{Weir_etal:1995a}
{Weir} N.,  {Fayyad} U.~M.,    {Djorgovski} S.,  1995, \aj, 109, 2401

\bibitem[\protect\citeauthoryear{{Wolf}, {Meisenheimer} \& {R{\"o}ser}}{{Wolf}
  et~al.}{2001}]{Wolf_etal:2001}
{Wolf} C.,  {Meisenheimer} K.,    {R{\"o}ser} H.-J.,  2001, \aap, 365, 660

\bibitem[\protect\citeauthoryear{Yasuda et~al.,}{Yasuda
  et~al.}{2001}]{Yasuda_etal:2001}
Yasuda N.,  et~al., 2001, \aj, 122, 1104

\bibitem[\protect\citeauthoryear{{York} et~al.,}{{York}
  et~al.}{2000}]{York_etal:2000}
{York} D.~G.,  et~al., 2000, \aj, 120, 1579

\end{thebibliography}

\bsp

\label{lastpage}

\end{document}